\documentclass[universe,review,accept,pdftex,oneauthor]{Definitions/mdpi} 
\firstpage{1} 
\pubvolume{11}
\issuenum{6}
\articlenumber{167}
\pubyear{2025}
\copyrightyear{2025}
\externaleditor{} % More than 1 editor, please add `` and '' before the last editor name
\datereceived{6 March 2025} 
\daterevised{9 May 2025} % Comment out if no revised date
\dateaccepted{21 May 2025} 
\datepublished{24 May 2025} 
\hreflink{https://doi.org/10.3390/\linebreak universe11060167} % If needed use \linebreak
\pdfoutput=1 % Uncommented for upload to arXiv.org
\usepackage{soul}

\usepackage{graphicx}
\usepackage{natbib}
\usepackage[utf8]{inputenc}
\newcommand{\Mpc}{$h^{-1}$\thinspace Mpc}

\newcommand{\vmh}{h^{-1}\mathrm{Mpc} }

\def\nat{Nature}
\Title{Galaxy Superclusters and Their Complexes in the Cosmic Web}

\TitleCitation{Galaxy Superclusters and Their Complexes in the Cosmic Web}

\Author{{Maret Einasto} %MDPI: Please carefully check the accuracy of names and affiliations.
  \orcidA{}} % $^{1,\dagger,\ddagger}$\orcidA{}} 
\AuthorNames{Maret Einasto}

\AuthorCitation{{Einasto}, M.%MDPI: Please check all author names carefully. %ME comment: name is correct
}

\address[1]{Tartu Observatory, University of Tartu, Observatooriumi 1, 61602 T\~oravere, Estonia; %MDPI: please add the Specific address, %including city， postcode and country.
maret.einasto@ut.ee}

\abstract{The richest and largest structures in the cosmic web are galaxy superclusters, 
their complexes (associations of several almost connected very rich superclusters),
and planes. 
Superclusters represent a special environment where the evolution
of galaxies and galaxy groups and clusters differs from the evolution 
of these systems in a low-density environment. The richest galaxy clusters reside in
superclusters. 
The richest superclusters in the nearby Universe 
form a quasiregular pattern with 
the characteristic distance between superclusters  120--140~\Mpc.
Moreover,
  superclusters in the nearby Universe lie 
in two~huge perpendicular planes with the extent of several hundreds of megaparsecs, 
the Local Supercluster plane and the Dominant supercluster plane. 
The origin of these patterns in the supercluster distribution 
is not yet clear, and it is an open question whether the presence 
of such structures can be explained within the $\Lambda$CDM cosmological model.
This review presents a brief story of superclusters, their discovery, definitions,
main properties, and large-scale distribution. }


\begin{document}

\section{Introduction}

\begin{figure}[H]
%\isPreprints{\centering}{} % Only used for preprints
\includegraphics[width=11 cm]{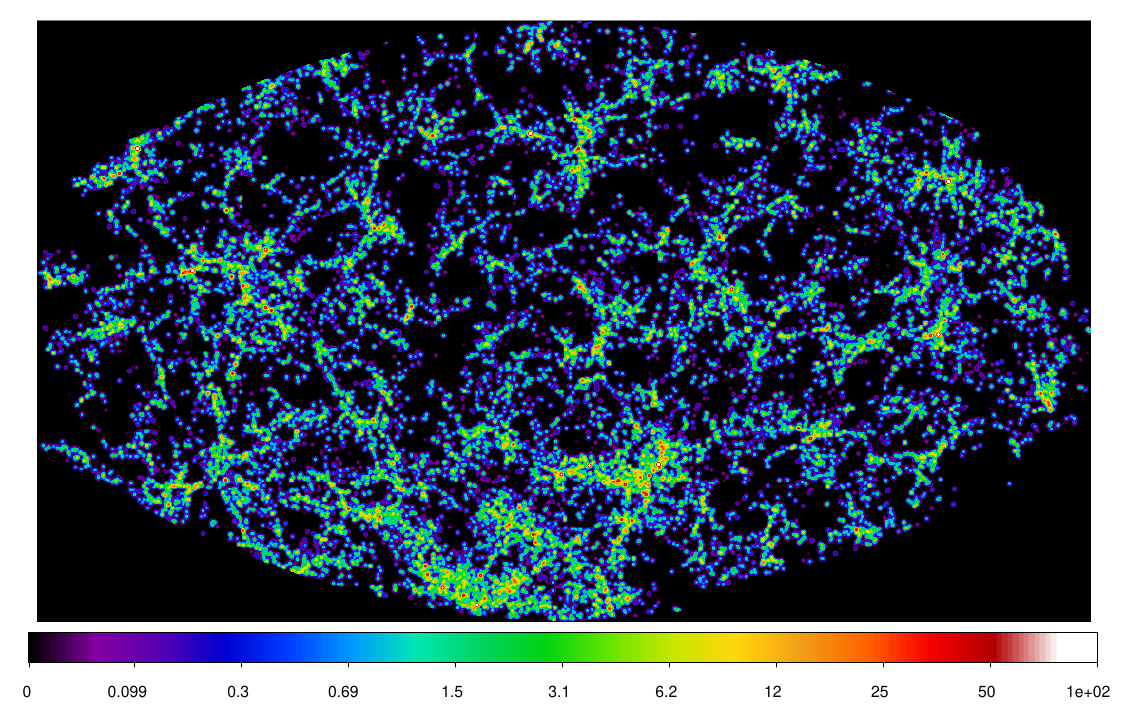}
\caption{{Slice} %MDPI: Please use scientific notations in the figure, e.g., “8 × 10³”, not “8E3”.
 of the density field from the Sloan Digital Sky Survey
at distance of 240~\Mpc\ with thickness of 10~\Mpc\ in the SDSS survey coordinates,
$\eta$ and $\lambda$.
At lower part of the figure the Sloan Great Wall is seen.
Figure by Jaan Einasto.  
\label{sdssslice}}
\end{figure}

One of the major achievements of the modern observational 
cosmology in the last century is the discovery that galaxies and their systems
form a web-like pattern now called the cosmic web, shown in Figure~\ref{sdssslice}
\citep{1978MNRAS.185..357J,
1980Natur.283...47E, 1983ARA&A..21..373O}. 
The web-like distribution of matter presents a huge variety of galaxy systems
from the galaxy clusters as nodes of the cosmic web, connected by elongated 
filaments of groups and single galaxies, and separated by huge  underdense regions with almost
no visible matter. 
The Cold Dark Matter cosmological model was accepted as a standard
model in the 1990 (see, for example, Critical Dialogues in Cosmology \citep{1997cdcp.book.....T}).
According to the standard Cold Dark Matter cosmological model with cosmological constant
($\Lambda$CDM) model, dark matter is the main matter component in the cosmic web,
alongside dark energy, and the formation and evolution 
of galaxies, galaxy groups, and clusters is governed by the dark
side---dark matter and dark energy.
Many methods have been developed to quantify the cosmic web and to study its
properties. One of the first and most common methods among them
is the two-point correlation function and its counterpart---power spectrum
\citep{1980lssu.book.....P, 2002sgd..book.....M}.
To describe the morphology of the cosmic web, higher-order statistics are used,
as N-point correlation functions, 
the wavelet analysis, clustering and percolation analysis, various morphological methods, Minkowski functionals,
Genus, Betti numbers, and shapefinders among them, Nth nearest neighbours, graph-based
methods, fractal analysis, spherical collapse, Gaussian mixture modelling, and others. A review of these methods can be found
in \citep{1980lssu.book.....P, 1996ApJ...465..499G, 1982fgn..book.....M, 2002sgd..book.....M,  
2017ApJ...845...55P, sah98, sss04, 2005EJASP2005...99S, 2005ApJ...634..744M,
2009LNP...665..291V, 2009LNP...665.....M, 2011MNRAS.416.2494P, 2020MNRAS.493.5972H}. 

To detect and classify various cosmic web elements, many special
methods have been developed, such as clustering analysis, various methods to determine cosmic filaments,
and so on \citep{2018MNRAS.473.1195L}.
Based on these methods, cosmic web elements can be divided into dense nodes, elongated filaments,
flattened sheets, and voids between them \citep{2013MNRAS.429.1286C, 2018MNRAS.473.1195L}. 
Among these methods, the percolation and clustering methods stand up as the first
statistical methods to determine superclusters.

High-density regions of the cosmic web, which embed galaxies,
groups, and clusters connected by filaments, are called galaxy superclusters.
With their rich inner structure,
superclusters as hosts of all cosmic web elements  can be considered miniature
Universes~\mbox{\citep{2021A&A...649A..51E, 2023ApJ...958...62S, 2024A&A...689A.332A}}. 
The richest superclusters may embed several tens of rich (Abell) clusters
and their sizes may reach 200~\Mpc, while poor superclusters
like our own, Local or Virgo supercluster, consist of only one rich cluster
surrounded by filaments of poorer groups, including the Local Group of galaxies, and having
an extent of approximately 20~\Mpc. Superclusters host the richest galaxy clusters and the brightest
galaxies (which are the brightest galaxies of rich clusters).
Opposite regions to superclusters are underdense regions between superclusters,
called cosmic voids. 
 Among well-known voids and supervoids, for example, are the Local void, the Bootes void, the Eridanus supervoid,
and others, with sizes exceeding 100~\Mpc\
\citep{1981ApJ...248L..57K, 2012ApJ...759L...7P, 2015MNRAS.450..288S, 2016MNRAS.462.1882K}. 
Superclusters host approximately $15$\% of all galaxies, and
they occupy 1\% of the total volume of the Universe, while voids fill 70--90\% of 
the total volume of the Universe \citep{2013MNRAS.429.1286C}.
In overdense regions (superclusters),  the growth of structures 
and the evolution of 
galaxies and galaxy systems are different  from those in cosmic underdense regions (voids)
\citep{2016MNRAS.458..394B, 2022A&A...668A..69E,  2024A&A...681A..91E, 2023Natur.619..269D}. 
Superclusters, as deep potential wells, act as attractors toward which
matter flows from their surrounding low-density regions~\citep{1988ApJ...329..519D, 1989ASSL..151..179B, 1990ApJ...364..370B,
1990Natur.343..726B, 2017ApJ...845...55P, 2023A&A...678A.176D, 2025arXiv250201308C}.

The presence of superclusters or clusters of clusters (second-order clusters)
in the distribution of galaxy clusters was noticed already 
by Harlow Shapley in 1930 \citep{1930BHarO.8749S}. This structure is now known 
as the Shapley supercluster, and it is 
the richest supercluster in the nearby Universe \citep{2024A&A...689A.332A}.
Our own Local Group of galaxies is located on the outskirts of the Local or Virgo
supercluster \citep{1953AJ.....58...30D, 1982ApJ...257..389T}.

The discovery of the cosmic web initiated the question on the co-evolution
of various elements of this pattern.
One of the first superclusters studied in detail is the Perseus-Pisces
supercluster, which is an elongated structure with the Perseus cluster as the richest galaxy cluster
in it (Abell cluster A~426), and with other clusters and groups
\citep{1978MNRAS.185..357J, 1980MNRAS.193..353E}.
These  studies  revealed correlated
alignments of the brightest cluster galaxies and galaxy groups and clusters
along the supercluster axis, as well as the morphological segregation of galaxies
in the supercluster. Early-type galaxies are located mostly in rich clusters
and along the main body of the supercluster, while late-type galaxies lie
in the outskirt regions---filaments and poor groups.
These findings suggest that galaxies, groups,
and superclusters evolve together \citep{1980MNRAS.193..353E}.
  
This review provides a brief overview of the studies of galaxy superclusters.
First, I present various definitions of superclusters, and 
supercluster catalogues determined based on different definitions and traced by 
different objects: galaxies, optical and X-ray clusters, and other objects.
Next, I discuss supercluster morphology, mass determination, and structure, 
biasing problems (relative distribution of visible and dark
matter in superclusters), fractal properties of superclusters,  
as well as the evolution of
superclusters in the cosmic web. 
Then, I describe superclusters as a special environment in which galaxies and
galaxy systems form and evolve. An  important part of this review is dedicated to
the largest systems in the cosmic web, supercluster complexes and planes,
and to the almost regular pattern detected in the supercluster distribution,
with the  120--140 $h^{-1}$Mpc scale between rich superclusters.
It has been suggested in the literature (e.g., \citep{1997Natur.385..112K})
that its origin is related to the Baryon Acoustic Oscillations. I will provide arguments
against this interpretation and in support of the idea that the origin of this pattern is related
to the dark matter perturbations in the very early Universe. 

The extreme cases of observed objects usually provide the
most stringent tests for theories; this motivates the need for
a detailed understanding of various properties of the richest superclusters,
their complexes, and planes. 
The properties of the high-density cores of superclusters have been proposed as a cosmological
probe to test the standard Cold Dark Matter cosmological model with cosmological constant
($\Lambda$CDM) model \citep{2021A&A...649A..51E, 2024PASA...41...78Z}.
Therefore, I discuss
whether the presence of such planes and the regular pattern in the supercluster
distribution  is compatible with the standard
$\Lambda$CDM model.

The early reviews of the first studies of superclusters were by  Jan Oort \citep{1983ARA&A..21..373O}
and by Neta Bahcall \citep{1996astro.ph.11148B}. 
The story of the discovery and present studies of the dark matter, cosmic web, and superclusters 
can be found in the book  ``Dark matter and cosmic web story'' by Jaan Einasto \citep{Einasto:2024aa}. 
In this book, several chapters are dedicated to the study of superclusters
in the cosmic web based on both simulations and observations.
In my review, I focus on observational studies. 
In the further text, I use the term the nearby Universe for the redshift up to
approximately $z \approx 0.1$. The local Universe  refers
to our close cosmic neighbourhood (mostly the Local, Virgo
supercluster) with redshifts $z < 0.02$.

%%%%%%%%%%%%%%%%%%%%%%%%%%%%%%%%%%%%%%%%%%

\section{Supercluster Definitions and Catalogues }
\label{sect:def}

Superclusters as the connected overdensity regions in the cosmic web have been defined 
on the basis of individual objects (typically optical or X-ray groups and/or clusters of galaxies) or
luminosity-density or velocity fields, calculated using galaxy \mbox{data \citep{1994MNRAS.269..301E, 
2006A&A...459L...1E, 2007A&A...462..397E, 2011MNRAS.415..964L, 2014Natur.513...71T, 2023A&A...678A.176D, 2023ApJ...958...62S, 2024ApJ...975..200C}.}
One of the most common methods to detect superclusters in the distribution of groups and
clusters of galaxies is {\it the cluster analysis or the Friend-of-Friend (FoF; percolation) method} \citep{1982Natur.300..407Z, 1982ApJ...257..423H}.
In this method, systems (in our case---superclusters) are determined using a certain neighbourhood
radius or linking length  which is used to link galaxies, groups, or clusters
together into systems. The neighbourhood of each galaxy group and/or cluster up to a certain neighbourhood radius
or linking 
length is searched for  neighbouring groups/clusters. 
Every group/cluster closer than the linking length to any 
other group or cluster  in the system  is considered a member of the system. If the linking length is small, then all clusters
 are isolated. With the increase in the linking length at first systems of clusters form at cluster pairs and
 triplets. 
With the further increase in linking length the number of systems 
and their richness (number of members)  increases. At a certain linking length, individual 
systems start to join into huge systems, which may penetrate the whole volume under study, and a
percolation occurs.
Therefore, it is important to find a linking length which separates individual superclusters before percolation.

When using {\it {the luminosity-density field}},  the  first step is to choose a proper smoothing length to calculate
the luminosity-density field in order to detect
superclusters. Next step is to choose the density level (usually in the units of the mean luminosity-density
of the sample) to define superclusters. 
For example,  {ref.} %MDPI: We added ``ref.'' before the reference at the beginnaing of the sentences, please check in the whole text.
\citep{2012A&A...539A..80L} determined the luminosity-density field using
the Sloan Digital Sky Survey MAIN galaxy data and smoothing kernel based on the $B_3$ spline function
with 8~\Mpc\ smoothing length to determine superclusters in the galaxy data (see also \citep{e07}
for kernel definitions).
In this review, the global luminosity-density determined in this way is 
denoted as $D8$.

\begin{figure}[H]
%\isPreprints{\centering}{} % Only used for preprints
\includegraphics[width=7 cm]{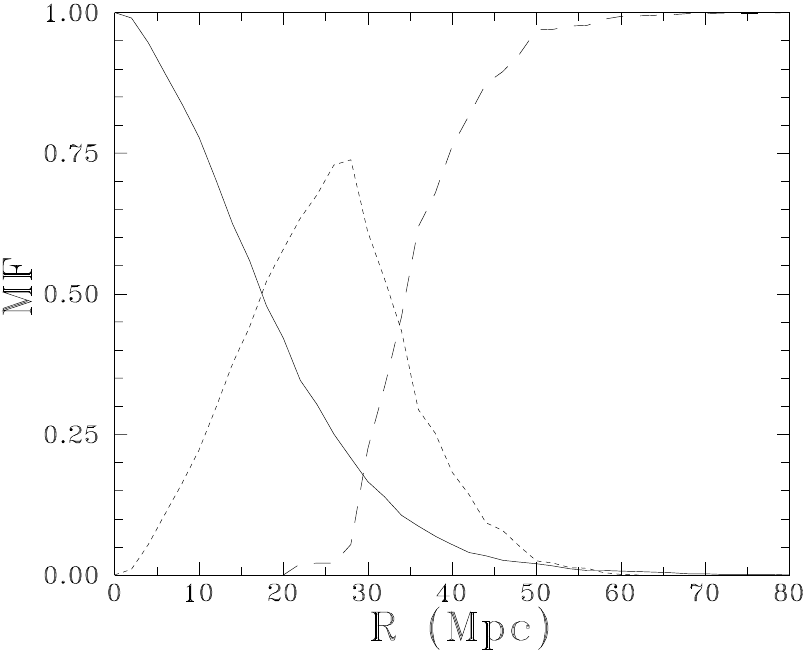}
\caption{Multiplicity functions  (which show the fraction of clusters in superclusters
of different richness) $MF$ of systems of Abell
clusters versus FoF neighbourhood radius $R$. The solid line shows the fraction of isolated
clusters. The short-dashed line shows the fraction of clusters in
medium-rich systems with a number of members from 2 to 31.  The dashed line
shows the fraction of clusters in very rich systems with at least 32
member clusters. Adapted from \citep{2001AJ....122.2222E}.  
\label{mfabell}}
\end{figure}

In the determination of superclusters, several criteria have been used to select the linking length or density 
level. Widely, these can be divided into three classes. First, superclusters have been defined as 
{\it {connected systems with linking length or density threshold before percolation}}, having, for example, the
maximum number of superclusters in the sample.
I illustrate this in Figure~\ref{mfabell},  which shows fractions of clusters
in superclusters of different richness (multiplicity function $MF$) 
versus the neighbourhood radius (linking length) $R$ 
for rich Abell clusters~\citep{2001AJ....122.2222E}. This figure shows that at small linking 
lengths all clusters are
isolated. With increasing neighbourhood radius, clusters are gathered at first into
systems of intermediate richness (with the number of clusters in superclusters or supercluster 
richness $2 \le N_{CL} \le
31$), and the fraction of single (isolated) clusters decreases. 
At larger radii, extremely large superclusters with multiplicity (number of clusters)
$N_{CL} \ge 32$ start to form. By further increasing the neighbourhood
radius, superclusters begin to merge into huge structures, and finally,
all clusters percolate and form a single system penetrating the whole
volume under study. Each linking length value divides objects under study
into high-density populations (connected systems), and low-density populations
\mbox{(isolated objects).}

\begin{figure}[H]
%\isPreprints{\centering}{} % Only used for preprints
\includegraphics[width=12 cm]{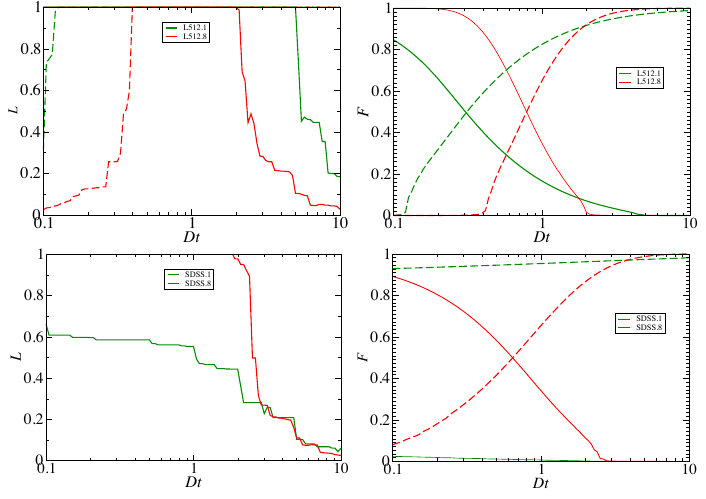}
\caption{Percolation functions of $\Lambda$CDM model of the box 
size 512~\Mpc~(top), and SDSS galaxy samples (bottom). 
Left panels show percolation length functions $L$ in unit of 
the box size, right panels filling factor functions $F$ in units 
box volume. Green lines correspond to the  smoothing length 1\Mpc, 
red lines correspond to the  smoothing length 8~\Mpc, solid lines represent clusters, and
dashed lines show voids. Figure by Jaan Einasto.  
\label{fill}}
\end{figure}

With percolation methods, it is possible to follow 
systems in over- and underdense regions, and to study their properties in detail,
both from simulations and observations.
\mbox{Figure~\ref{fill}} shows the lengths $L$ and filling factors $F$ of systems as a function of
threshold density $D_t$. 
The upper panels of \mbox{Figure~\ref{fill}} show these properties for 
$\Lambda$CDM simulation with 
resolution $N_{\mathrm{grid}} = 512$ and $N_{\mathrm{part}} = 
N_{\mathrm{grid}}^3$ particles, and with the cube size \mbox{$L_0=512$~\Mpc. }
The lower panels show distributions for galaxy groups from the SDSS DR8.
For superclusters (connected cosmic web elements determined
with the smoothing length 8~\Mpc)  the filling factor  $F_{scl} = V_{scl}/V_{tot}$,
where $V_{scl}$ denotes the volume occupied by superclusters, and $V_{tot}$ denotes
the total volume of the sample. Filling factors of other cosmic web elements 
are defined in the same way.
Cluster sizes are expressed 
in units of the sample size, $L_0$ (the effective side length in the 
case of the SDSS sample).
Figure~\ref{fill} shows the lengths and volumes of the largest clusters and voids 
(percolation functions) as functions of the threshold density, $D_t$.

As seen in this figure, 
with the increase in the threshold
density $D_t$, the volume occupied by voids increases, and the volume covered by 
large systems decreases. At the threshold density  $D_t \approx 5$ superclusters
occupy approximately 1\% of the total volume, as shown also by \citet{2013MNRAS.429.1286C}.
At a very high threshold density, there exist only a few high-density 
regions---peaks of the density field as ordinary clusters of galaxies.  
These peaks are isolated from each other and cover a small volume (filling factor) in space. 
The percolation functions of superclusters  for $\Lambda$CDM model 
and SDSS samples are \mbox{rather similar.}

One can also note that the percolation functions of voids for the $\Lambda$CDM model and SDSS samples 
are very different. At a very low threshold density, void sizes of 
the  $\Lambda$CDM model are small; they form isolated bubbles 
inside the large over-density cluster, and the filling factor of 
the largest void is very small. Void bubbles are separated from 
each other by dark matter sheets.  Some sheets have tunnels that permit the 
formation of some larger connected voids.  With increasing threshold 
density, the role of tunnels rapidly increases, and tunnels join neighbouring voids.  
At a certain threshold density, the largest void is percolating, 
but still not filling a large fraction of the volume. 
The major difference between models and observations is the absence in 
SDSS samples of the fine structure of voids. 
At all smoothing lengths, SDSS voids are percolating, and the percolation threshold 
density is not defined.  For small smoothing lengths, the percolating 
SDSS void is the only void.

\begin{figure}[H]
%\isPreprints{\centering}{} % Only used for preprints
\includegraphics[width=11 cm]{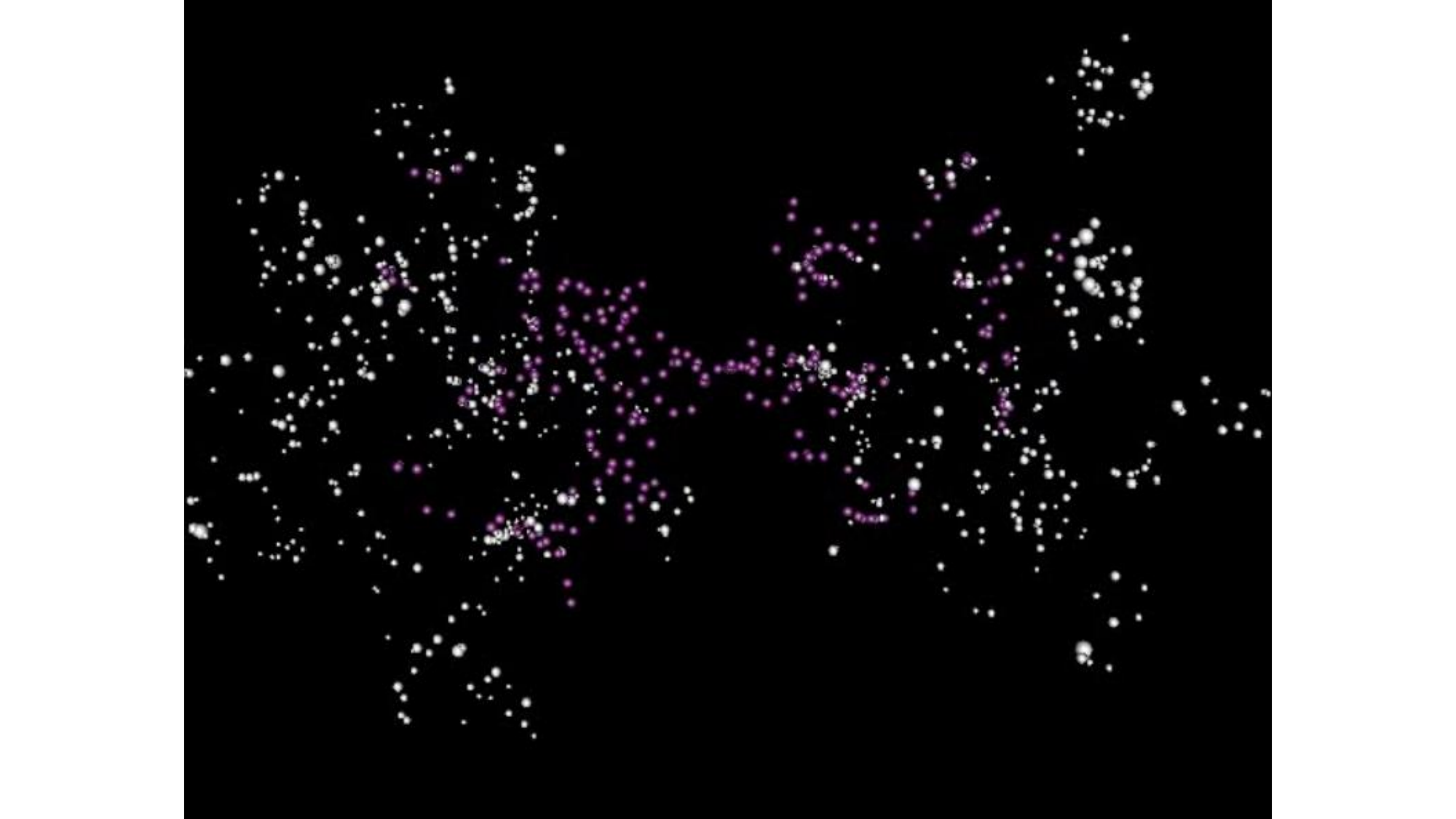}
\caption{The 3D distribution of rich (Abell) clusters in superclusters (white spheres)
in supergalactic coordinates. Violet spheres show the location of X-ray clusters
in superclusters. Note the signature of the Local supercluster plane in the centre
of the figure (Section~\ref{sect:planes}).
Figure by ME, based on data from~\citep{2001AJ....122.2222E}.  
\label{abellx}}
\end{figure}

 In order to obtain superclusters as the largest
still relatively isolated systems a neighbourhood radius should be
smaller than the percolation radius. In \citep{2001AJ....122.2222E},
who used data on rich (Abell) clusters of galaxies to detect high-density regions---superclusters in the cluster distribution,
 the  linking length 24~\Mpc\
was selected to define superclusters. %MDPI: Please confirm if the bold formatting is necessary; if not, please remove it. The following highlights are the same.
This value is close to the Poisson radius of Abell clusters (radius of the sphere
that contains one cluster, 
$R_P = 3V/(4\pi N)^{1/3}$; where $V$ is the volume of the sample, and $N$ 
is the number of clusters in the sample. 
$R_{P(Abell)} = 26$~\Mpc). At this linking length, the size of the largest 
supercluster is approximately 100--150~\Mpc. 
The first
catalogues of superclusters of rich (Abell) clusters were compiled in this way
\citep{1989ApJ...347..610W, 1994MNRAS.269..301E, 1997A&AS..123..119E, 2001AJ....122.2222E, 2014MNRAS.445.4073C}. 
Ref. \citep{1993ApJ...407..470Z} presented supercluster catalogues based on Abell clusters for several linking lengths. 
Figure~\ref{abellx} shows the 3D distribution of Abell clusters and X-ray clusters
in superclusters, based on data from~\citep{2001AJ....122.2222E}.
This criterion was adopted in a recent catalogue
of superclusters in redshift interval $0.05 < z < 0.42$ by \citep{2023ApJ...958...62S}.

Recently, a huge structure was detected, using data on X-ray clusters, and nicknamed
  {\it  the Quipu superstructure} \citep{2025arXiv250119236B}.
Cluster distribution in this structure is rather sparse, and these were linked 
together applying the Friend-of-Friend method with a large linking length.
Therefore, the authors call it a ``superstructure'', not a supercluster.
The Quipu superstructure embeds several poor superclusters of Abell clusters,
determined in the supercluster catalogue by
\citep{2001AJ....122.2222E}, and a long filament that connects these superclusters.
This is owing to the use of a larger linking length than in \citep{2001AJ....122.2222E}
to link clusters together into \mbox{a system. }

%%%

\begin{figure}[H]
%\isPreprints{\centering}{} % Only used for preprints
\includegraphics[width=11 cm]{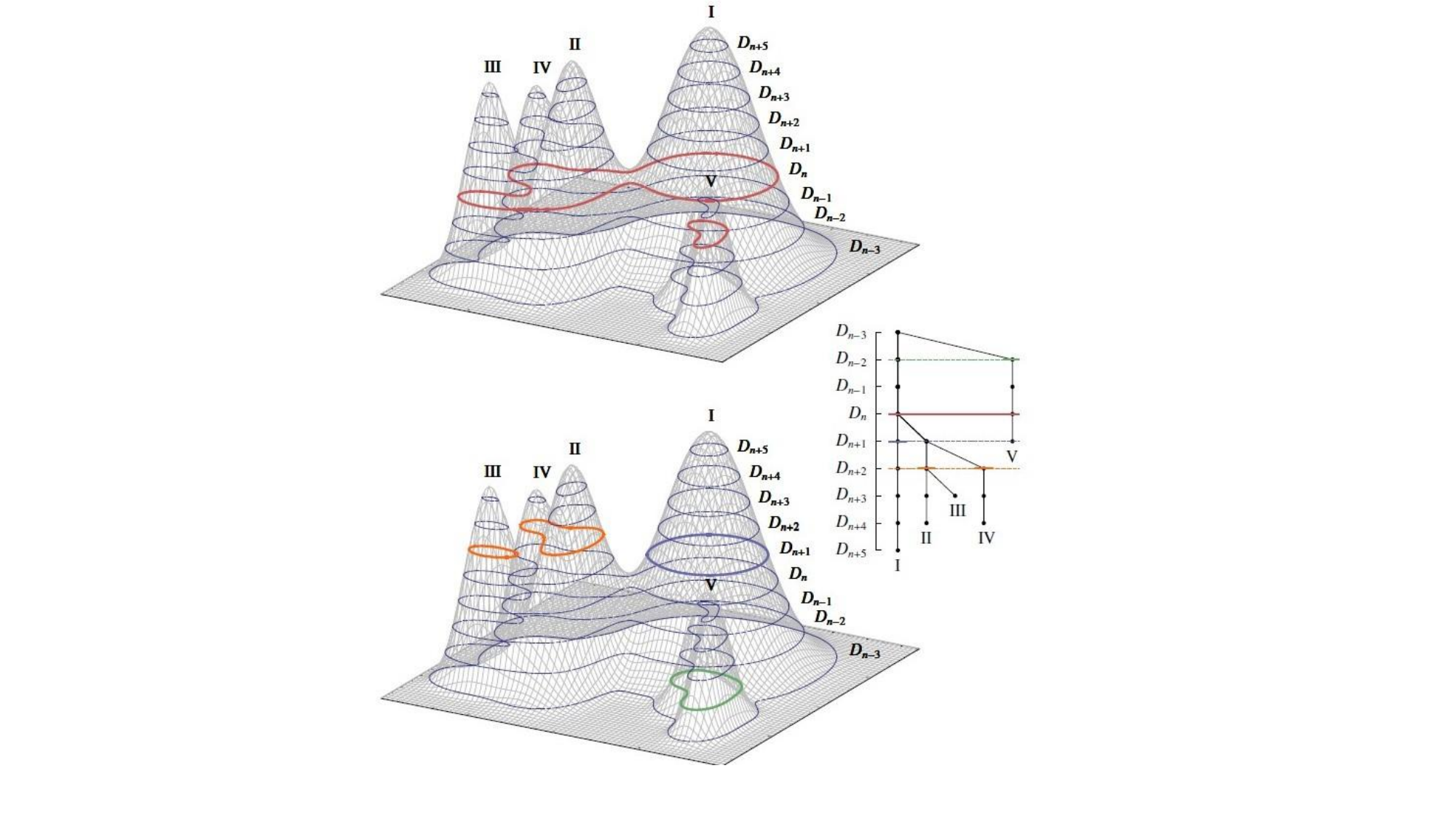}
\caption{{Density} %MDPI: Please check if the explanation if the different color and letters should be added.
 field structures (thick red, orange, and dark blue coloured contours) detected with a single
fixed density level $D_n$ (upper panel) or with various individually derived thresholds (lower
panel). Thin contours denote all available isodensity cuts of the density field. The corresponding
tree structure of the density field objects is drawn in the middle. Adapted from \citep{Juhan_thesis}.  
\label{denlevels}}
\end{figure}

In the case of the luminosity-density field, the largest
superclusters have approximately  the same size, 100--150~\Mpc, when applying the threshold density level $D = 5$
(in units of mean luminosity density) for the SDSS 
MAIN galaxy sample (see \citep{2012A&A...539A..80L},
and also~\citep{2011ApJ...736...51E} in the case of the Sloan Great Wall superclusters).
Ref. \citep{2012A&A...539A..80L} applied  two different criteria
for the density level to determine superclusters in the galaxy distribution, 
fixed and adaptive density 
levels. In catalogues with a fixed density level, the criteria for the 
density level can be searched for in the same way as in the case of linking length,
i.e., the threshold density for the selection of superclusters should be selected so
that it separates individual superclusters before percolation
\citep{2011A&A...532A...5E, 2012A&A...539A..80L}. In addition, ref. \citep{2012A&A...539A..80L} 
searched for each supercluster
its individual characteristic density level at which a given supercluster is separated 
from the neighbouring superclusters. The resulting catalogue of superclusters is called
adaptive catalogue. 
The choice of superclusters with fixed and adaptive density levels is illustrated
in \mbox{Figure~\ref{denlevels}~\citep{2012A&A...539A..80L, Juhan_thesis}}.
Typically, superclusters determined using these criteria are not gravitationally bound.

Another criterion is to search for the largest structures that will survive the cosmic expansion in an
acceleratingly expanding Universe, and will remain bound also in the future ({\it future collapsing 
superclusters or superstes--clusters}) \citep{2011MNRAS.415..964L, 2013MNRAS.429.3272C, 
2015A&A...575L..14C, 2024ApJ...975..200C}. 
%EE: Please check that the intended meaning has been retained %ME changed to superstes--clusters
 The first  catalogue of superclusters, 
in which superclusters, determined using the luminosity-density field, which was applied to 
the SDSS MAIN galaxy data, were defined as structures that evolve into isolated, virialised structures in the future,
was by \citep{2011MNRAS.415..964L}.
% spherical collapse based criterion
Using this criterion, ref. \citep{2024ApJ...975..200C} compiled a catalogue of superclusters in the 
redshift interval $z$ = 0.5--0.9, 
based on  the CAMIRA cluster sample constructed
using the Subaru Hyper Suprime-Cam survey data. Although individual superclusters have also been detected at
higher redshifts, this is at present the highest redshift interval at which supercluster catalogues have been determined.

\begin{figure}[H]
%\isPreprints{\centering}{} % Only used for preprints
\includegraphics[width=11 cm]{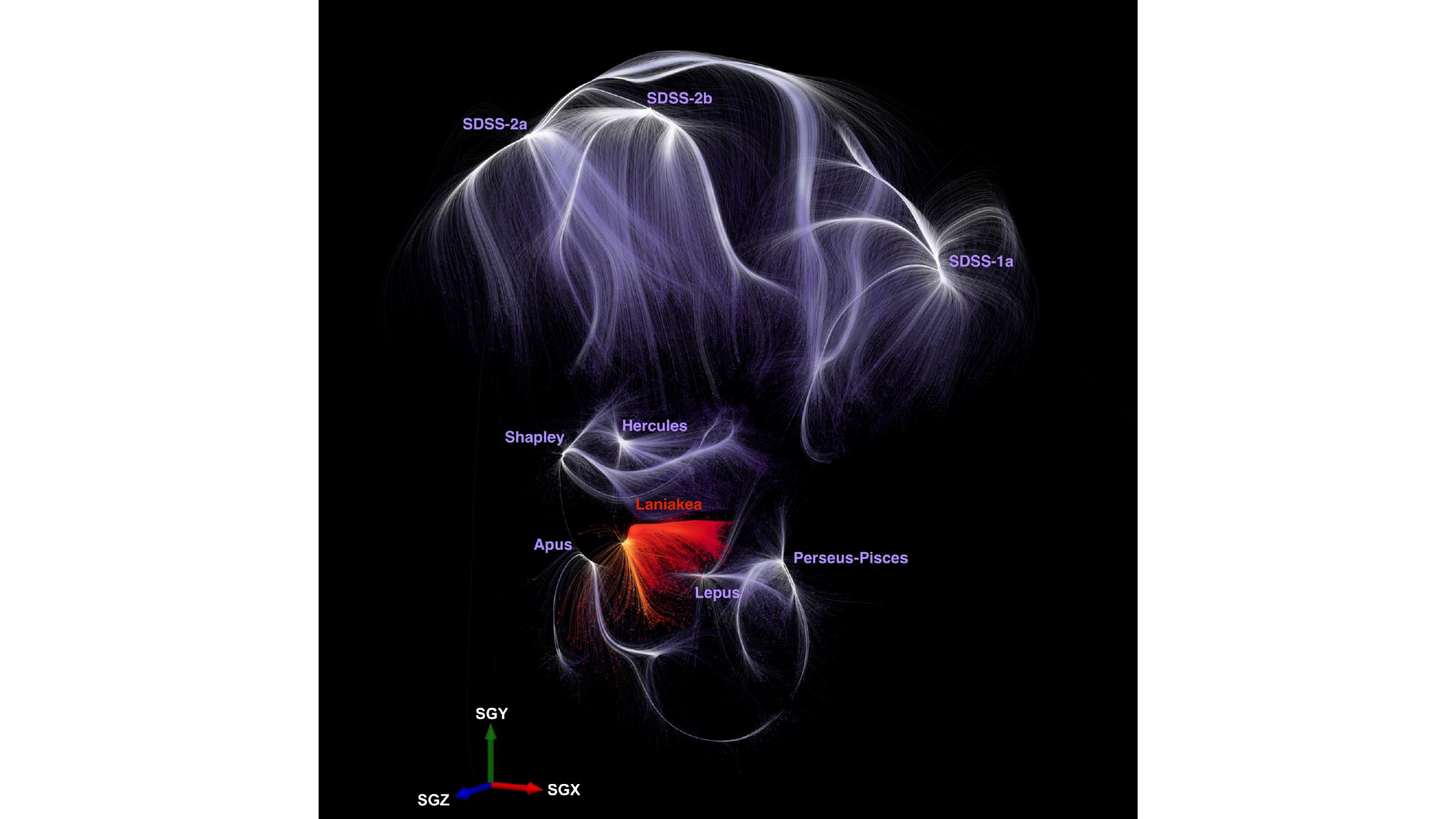}
\caption{Visualization of the BoAs of Laniakea 
(in red) and the  Apus, the Hercules, the Lepus, the Perseus-Pisces, 
and the Shapley superclusters, as well as the BoAs in 
the SDSS area (in purple). The basins shown are obtained 
from the ungrouped CosmicFlows-4 (CF4) velocity field. 
Streamlines 
are shown for each galaxy with the same colour code. 
The gradient of colour is related to the density of streamlines. 
For reference, the three supergalactic cartesian orientation axes SGX, SGY, SGZ 
are drawn on the bottom left, represented, respectively, by red, green, and blue arrows 
of size 50 Mpc $h^{-1}$. 
Figure by Alexandra Dupuy and Helene Courtois \citep{2023A&A...678A.176D}.  
\label{baos}}
\end{figure}

Third, using the velocity field of galaxies, 
large-scale structures in the cosmic web---superclusters and voids---can be defined as regions in space 
where galaxy velocities are directed inward
or outward under gravity. 
%EE: Please check that the intended meaning has been retained % This is OK
Galaxy superclusters are then defined as 
{\it converging peculiar velocity
field regions}---regions, where, on average, peculiar velocities of galaxies are pointing towards
the central high-density regions 
(basins of attraction or BoA) 
\citep{2014Natur.513...71T, 2015ApJ...812...17P,
2019MNRAS.489L...1D, 2023A&A...678A.176D}. 
Superclusters as BoAs are volumes in space in which the velocities of galaxies and
galaxy groups are, on average, pointing toward a central attractor. 
The central attractor of a supercluster may be the richest cluster in it.
In the case of BoAs with large, typically elongated, central high-density regions that embed several
rich clusters, the central attractor may lie on the axis or filament between
clusters  \citep{2023A&A...678A.176D}.
In the case of large BoAs, such as the Laniakea supercluster \citep{2014Natur.513...71T}, 
only the central,  highest-density parts 
may collapse in the future \citep{2015A&A...575L..14C}, while smaller BoA superclusters, such as, for example,
the Arrowhead supercluster \citep{2015ApJ...812...17P},
may fully collapse during future evolution.
As an example, Figure~\ref{baos} presents the BoAs of superclusters in the nearby Universe from \citep{2023A&A...678A.176D}.

Opposite to BoAs, basins of repulsions (BoRs) are defined as regions in which peculiar velocities of
galaxies are directed outwards \citep{2023A&A...678A.176D}.
Simulations show that the whole space can be divided between BoAs \citep{2021MNRAS.500L..32P}. 
 The borders of BoAs (watershed regions) are regions where the values of
velocities are the lowest, or,
in other words, these are surfaces where velocity flows (streamlines) diverge, 
pointing at one side to the one BoA central attractor, and 
pointing on the other side to the central attractor of another BoA. 
Conventional superclusters form central high-density regions of BoAs. 
In order to keep the traditional definition of superclusters as the high-density regions of the cosmic
web, ref. \citep{2019A&A...623A..97E} proposed to call low-density regions around superclusters 
in the BoAs supercluster cocoons. Simulations show velocity change at supercluster
borders within BoAs \citep{2024A&A...687A..85S}.

Supercluster catalogues have also been generated using a different approach.
For example, catalogues of cosmic voids have been generated using the watershed transform
algorithm, and using the inversion of this algorithm, catalogues of overdensities or
``superclusters'' have been generated and made available \citep{2014MNRAS.440.1248N}.

As already mentioned, first supercluster catalogues were compiled using data on rich optical (Abell) clusters 
of galaxies. Refs. \citep{1994MNRAS.269..301E, 2001AJ....122.2222E} also included X-ray clusters into 
their supercluster catalogues. They noted
that X-ray clusters tend to be more strongly clustered than optical clusters. First catalogues
based on  X-ray clusters only based on ROSAT data were compiled by  \citep{2013MNRAS.429.3272C, 2021A&A...656A.144B}.
Ref. \citep{2024A&A...683A.130L} compiled a catalogue of superclusters of X-ray-selected superclusters 
identified in the first SRG/eROSITA
survey in the western Galactic hemisphere up to the redshift $z = 0.8$.
We also mention that the Friend-of-Friend method was applied to compile  quasar system catalogues
\citep{2014A&A...568A..46E, 2015JKAS...48...75P}.
The web links of some supercluster catalogues mentioned in the text and available online 
are given in %MDPI: We revised the footnote to note, please check if the word should be revised.
%MDPI: Please add the access date for all links (format: Date Month Year), e.g., accessed on 1 January 2020.
 {\it the Note}~\citep{2001AJ....122.2222E, 2012A&A...539A..80L, 2014MNRAS.445.4073C, 
2020A&A...637A..31S, 2021A&A...656A.144B, 2024A&A...683A.130L, 2023ApJ...958...62S, 
2024ApJ...975..200C}\endnote{\url{https://vizier.cds.unistra.fr/viz-bin/VizieR?-source=J/AJ/122/2222}, inserted into VizieR on 22 July 2002;
\url{https://cdsarc.cds.unistra.fr/viz-bin/cat/J/A+A/539/A80}, inserted into VizieR on 27 February 2012; 
\url{https://cdsarc.cds.unistra.fr/viz-bin/cat/J/MNRAS/445/4073}, inserted into VizieR on 21 October 2015;  
\url{https://cdsarc.cds.unistra.fr/viz-bin/cat/J/A+A/637/A31}, inserted into VizieR on 08 May 2020;  
\url{https://cdsarc.cds.unistra.fr/viz-bin/cat/J/A+A/656/A144}, inserted into VizieR on 20 April 2022;  
\url{https://cdsarc.cds.unistra.fr/viz-bin/cat/J/A+A/683/A130}, inserted into VizieR on 13 March 2024; 
\url{https://content.cld.iop.org/journals/0004-637X/958/1/62/revision1/apjacfaebt2_mrt.txt}, DOI 10.3847/1538-4357/acfaeb;
\url{https://content.cld.iop.org/journals/0004-637X/975/2/200/revision1/apjad76adt1_mrt.txt, DOI: 10.3847/1538-4357/ad76ad}.
}.

\section{Supercluster Morphology}
\label{sect:morph}

Superclusters can be characterised using several properties, including their shapes, sizes,
masses, and others. I start the description of superclusters from the analysis of their shapes and sizes.
The overall shape of superclusters was analysed, approximating their shapes
with triaxial ellipsoids \citep{1989ApJ...347..610W, 1998A&A...336...35J}. These studies show
 that superclusters are elongated;
even the most round superclusters  have the ratio of their shortest and longest
axis $a/c < 0.5$. One interesting result of these studies is the finding
that the most elongated,
thin superclusters were also the most perpendicular to the line-of-sight \citep{1998A&A...336...35J}. 
The richest supercluster
in the Sloan Great Wall complex, supercluster SCl~126, is
among them \citep{1997A&AS..123..119E}.  The authors interpreted this as a signature 
of  the large-scale peculiar motions 
toward the supercluster central axis, which makes the supercluster appear thinner than it actually is;
an evidence of the so-called``Kaiser effect'' \citep{1984ApJ...284L...9K}.

A more detailed analysis of the shapes and inner structure of superclusters is provided by the application
of Minkowski functionals (MFs) and so-called shapefinders calculated on the basis of MFs 
\citep{kbp02, e07, 2011MNRAS.411.1716C}.
Four MFs characterise morphology and topology of
isodensity contours, which embed superclusters. For a given surface, MFs are as follows:
\begin{enumerate}
\item The first Minkowski functional $V_{0}$ is the enclosed volume V; 
\item \textls[-15]{The second Minkowski functional
 $V_{1}$ is proportional to the area of the surface 
$S$, {namely,} %MDPI: Please confirm if the variables in equations should be united (italic, bold, subscript/superscript), please check in the whole text.
}
\begin{linenomath}
    \begin{equation} 
        V_1 = {1\over6} S;
    \end{equation}
\end{linenomath}
\item The third Minkowski functional
 $V_{2}$ is proportional to the integrated mean curvature~C, 
\begin{linenomath}
\begin{equation} 
        V_2 = \frac1{3\pi} C, \quad 
        C=\frac12\int_S\left(\frac1{R_1}+\frac1{R_2}\right)\,dS,
\end{equation}
\end{linenomath}
where $R_1$ and $R_2$ are the two local principal radii of curvature.
\item The fourth Minkowski functional
 $V_{3}$ is proportional to the integrated Gaussian 
curvature (or Euler characteristic) $\chi$, 
\begin{linenomath}
  \begin{equation}
  V_3=\frac12\chi,\quad 
  \chi = \frac1{2\pi}\int_S\left(\frac1{R_1R_2}\right)dS.
  \end{equation}
\end{linenomath}
\end{enumerate}

The Euler characteristic 
is related to the genus, $G$
\begin{eqnarray}
\label{genus}
   G&=&1-V_3.  
  \end{eqnarray}

It gives %MDPI: Please confirm if indent should be added at the beginning of the senternce after the equation, please check the whole text.
 the number of isolated clumps and voids in the enclosed volume:
\begin{linenomath}
\begin{equation}
\label{v3}
V_3=N_{\mbox{clumps}} + N_{\mbox{bubbles}} - N_{\mbox{tunnels}}.
\end{equation}
\end{linenomath}
On the basis of MFs, one can define the so-called shapefinders,
which characterise the thickness, breadth, and length of an object: 
$H_1=3V/S$ (thickness),
$H_2=S/C$ (breadth), and \mbox{$H_3=C/4\pi$} (length). These quantities have
dimensions of length, and these are normalised to give $H_i=R$ for a sphere
of radius $R$.  For a convex surface, the shapefinders $H_i$ follow
the inequalities $H_1\leq H_2\leq H_3$.  Prolate ellipsoids (pancakes)
are characterised by $H_1 << H_2 \approx H_3$, while oblate ellipsoids
(filaments) are described by $H_1 \approx H_2 << H_3$.
 Additionally, Sahni et al. \cite{sah98} defined  two dimensionless
shapefinders $K_1$ (planarity) and $K_2$ (filamentarity): 
$K_1 = (H_2 - H_1)/(H_2 + H_1)$ and $K_2 = (H_3 -
H_2)/(H_3 + H_2)$ \citep{sah98, sss04}.
The overall morphology of a supercluster can be described by
the shapefinders $K1$ and $K2$, and their
ratio, $K1/K2$ (the shape parameter).

MFs and Genus have been applied to study the topology of the whole cosmic web, and to
compare it with the Gaussian random fields, as well as to study the properties 
of the Cosmic Microwave Background \citep{1996ApJ...465..499G, 1996MNRAS.281L..82C, 
2019MNRAS.485.4167P, 2021ApJ...907...75A}.
MFs and shapefinders have been used to study the integrated shapes of a set of superclusters from
mock catalogues, simulating data from the Sloan Digital Sky Survey (SDSS), and also to analyse morphological
properties of the cosmic web, from different cosmologies \citep{2004MNRAS.354..332S,
2003MNRAS.343...22S}. These studies showed that superclusters are elongated, dominantly
with filament morphology, with length up to approximately 100~\Mpc. 
These studies also suggested that geometrical properties
of the cosmic web are somewhat different in different cosmological models.
 
Ref. \citep{e07} applied MFs and shapefinders to characterise a small sample of individual richest superclusters
in detail, defined using the 2degree Field (2dF) galaxy data and data from the Millennium simulations. 
As a next step in the studies of supercluster morphology,
ref.~\citep{2011A&A...532A...5E} analysed the morphological properties of 36 individual rich superclusters determined by applying the 
SDSS galaxy data. Further, refs. \citep{Juhan_thesis,2023MNRAS.521.4712B} applied the same methodology
to study the morphology of a large sample of superclusters from SDSS.

\begin{figure}[H]
%\isPreprints{\centering}{} % Only used for preprints
\includegraphics[width=5 cm]{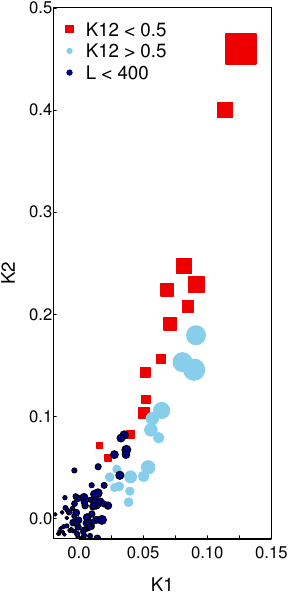}
\caption{Shapefinder's $K1$-$K2$ plane for superclusters. 
The size of symbols is proportional to the
diameters of superclusters. Red squares denote
high-luminosity superclusters with 
the luminosity $L_{scl} \geq 4 \times10^{12}~h^{-2} L_\odot$ and the
shape parameter $K1/K2 < 0.5$
(more elongated superclusters);
blue circles denote high-luminosity superclusters with 
the luminosity $L_{scl} \geq 4 \times10^{12}~h^{-2} L_\odot$ and the
shape parameter $K1/K2 > 0.5$
(less elongated superclusters); and
navy circles denote low-luminosity superclusters with $L_{scl} < 4 \times10^{12}~h^{-2} L_\odot$.
Adapted from  \citep{2011A&A...535A..36E}.  
\label{k12}}
\end{figure}

The studies of supercluster morphology  showed that poor and very poor superclusters are mostly spherical.
The richest superclusters are elongated; the limit between rich and poor superclusters
is approximately at supercluster sizes $D_{scl} \approx 20$~\Mpc, 
luminosities $L_{scl} = 4 \times10^{12}~h^{-2} L_\odot$,
and masses approximately 
$M \approx 5\times~10^{15}M_\odot$ \citep{2011A&A...535A..36E, 2023MNRAS.521.4712B, 2023ApJ...958...62S}.
According to their shape parameter $K12 = K1/K2$,  rich superclusters 
can be divided into less elongated and more elongated superclusters.
This is shown in Figure~\ref{k12}, which presents the shapefinders $K1$-$K2$ plane for superclusters
of different luminosity \citep{2011A&A...535A..36E}. Rich, more elongated superclusters
with $K12 < 0.5$ are richer and bigger than less elongated superclusters with $K12 > 0.5$.
The sizes of the largest superclusters may exceed 150~\Mpc\ \citep{2023ApJ...958...62S}.
There are only a few rich superclusters with a quasi-spherical shape \citep{2022A&A...668A..37H}.
The fourth MF characterises the inner structure of superclusters
(their clumpiness). Poor superclusters have typically small clumpiness, while rich superclusters
with complicated inner structure have high clumpiness values~\citep{2011A&A...535A..36E, 2023MNRAS.521.4712B}.

The physical
and morphological properties of superclusters are correlated: richer superclusters are more elongated,
and they have higher clumpiness values than poor superclusters. This enables us
to apply the Principal Component Analysis (PCA) to derive the scaling relations for
superclusters and to determine the fundamental plane of superclusters~\citep{2011A&A...535A..36E}.
This analysis showed that superclusters can be characterised with a small
number of parameters, as their luminosity, which is strongly correlated with supercluster mass
and size, and shape parameters (shapefinders and clumpiness). 
Correlation between physical and morphological parameters has been used 
to estimate supercluster masses (see Section~\ref{sect:mass} and \citep{2011A&A...535A..36E, 2017A&A...603A...5E}).

In the shapefinder's space, superclusters form a curve called the 
morphological signature, characteristic to multibranching, elongated objects \citep{e07, 2011A&A...532A...5E}.
At high densities within superclusters, 
the morphological signature of superclusters changes rapidly \citep{e07, 2011A&A...532A...5E}.
This property, together with the analysis of the density distribution
in superclusters have been used to define high-density cores of superclusters---high-density regions which may embed several rich galaxy clusters
(Section~\ref{sect:cores} and
\citep{2007A&A...464..815E}). If superclusters have been defined
as future collapsing systems, then such superclusters may correspond to 
high-density cores of superclusters defined using linking length or density level
near percolation \citep{2016A&A...595A..70E}.

One interesting result of the studies of supercluster morphology is that, according to their
inner structure, superclusters can be divided into filament-type and spider-type superclusters
\citep{e07,2011A&A...532A...5E}.
Filament- and multibranching filament-type superclusters have rather simple inner structure
with only a few galaxy filaments connecting galaxy clusters in the supercluster.
The richest supercluster in the Sloan Great Wall is an example of this type of supercluster.
In spider- and multispider-type superclusters, clusters are connected by many filaments.
The second richest supercluster in the Sloan Great Wall is of the multispider type~\citep{2011A&A...532A...5E}.
Superclusters with a
similar overall shape may have different fine structure \citep{2011A&A...535A..36E}.

We can ask the question whether the supercluster morphology and its large-scale
environment (location in the cosmic web) are related. This question was addressed in 
\citep{2011A&A...532A...5E}.
They analysed the location of superclusters, for which they determined inner morphology,
in the cosmic web. Most interestingly, this analysis showed that rich superclusters
at redshifts approximately $z$ = 0.07--0.08 form three chains of rich superclusters, separated
by voids. At the intersection of these chains lies the Corona Borealis supercluster, which is of 
multispider type. One chain is formed by superclusters in the Sloan Great Wall, which
are of multibranching filament type and multispider type. Superclusters in other
chains are not as rich as the Sloan Great Wall superclusters. These are the multispider-type
Bootes supercluster
and the filament-type Ursa Major supercluster, among others. Ref.~\citep{2011A&A...532A...5E} concluded that,
as the number of superclusters in their analysis was relatively small,
their study did not give a definite answer to the question
whether there is a connection between the morphology of superclusters
and their large-scale distribution. 
%%%%%%%%%%%%%%%%%%%%%%%%%%

The supercluster morphology can be used as a test for cosmological models
\citep{kbp02, e07}. 
Ref.~\citep{2007JCAP...10..016H} demonstrated that during the future evolution,
the structures in the cosmic web will sharpen. This affects the shape of superclusters,
which will  become more compact and round \citep{2009MNRAS.399...97A}.
Ref.~\citep{e07} showed that the morphology of observed superclusters
is not yet fully recovered among simulated superclusters.

\section{Supercluster Masses}
\label{sect:mass}

One of the most important characteristics of superclusters is their {\it mass}. Superclusters
represent overdensity regions of the cosmic web, where most of the
mass comes from the dark matter. Observationally, this mass is traced by visible structures such as galaxies,
groups, and clusters, connected by filaments, and intercluster gas. This introduces a bias in supercluster
masses and the estimation of 
supercluster masses may be challenging.
Therefore, this section presents the estimations of supercluster masses based on observations,
and the next section (Section~\ref{sect:bias}) is dedicated to the discussion of biases 
in the supercluster masses due to the dark matter content of superclusters, and how well this is traced by
visible structures in superclusters. 

Superclusters are high-density regions in the cosmic web, which embed galaxy groups and clusters,
as well as single galaxies, which do not belong to any detectable group (within observational limitations
of a particular observational data used to determine superclusters), and gas. Galaxy clusters in
superclusters are connected by filaments of poor groups, single galaxies, and gas, and the mass
in filaments is taken into account when considering masses of these structures.
Therefore, the mass of superclusters is the sum of the masses of these elements:
\begin{linenomath}
\begin{equation}
\label{mscl}
M_{scl}= \sum M_{\mbox{gr,cl}} + \sum M_{\mbox{single}} + M_{\mbox{gas}} + M_{\mbox{DM(additional)}},
\end{equation}
\end{linenomath}
where $\sum M_{\mbox{gr,cl}}$ denotes the total mass in groups and clusters in a supercluster,
$\sum M_{\mbox{single}}$ is the sum of masses of single galaxies, and $M_{\mbox{gas}}$ represents the
total mass of gas in a supercluster. With $M_{\mbox{DM(additional)}}$, we denote possible additional
dark matter in superclusters, which is not traced by visible structures in a 
supercluster---galaxies, groups, clusters, and gas. 
This will be discussed in the next section.

In these studies, which define superclusters on the basis of group and cluster data,
one common method to find supercluster masses
is the calculation of the sum of the masses of their member groups and clusters.
The catalogues of superclusters by \citep{2023ApJ...958...62S, 2024A&A...683A.130L}
provide masses of superclusters found in this way.

The supercluster catalogue by \citep{2023ApJ...958...62S} is based 
on galaxy groups from \citep{Wen12}. This catalogue presents masses of groups 
as found in \citep{Wen15}, $M_{500}$, which by \citep{2023ApJ...958...62S}
are converted to virial mass $M_{200}$ as follows:
$M_{200} = \frac{200}{500} \left(\frac{R_{200}}{R_{500}}\right)^{3} M_{500}$
(here mass $M_{500}$ and radius $R_{500}$, and mass $M_{200}$ and radius $R_{200}$
denote mass and radius, within which 
the mean density is $500$ and $200$ times the critical density of the universe).
For an NFW profile with a concentration parameter of about 4--8, 
$R_{500} \approx 0.65 \times R_{200}$ \citep{Ettori09}.
To obtain the bound halo mass for groups (including dark matter in groups), 
this mass is scaled: $M_{bound} \sim 2.2 \times M_{200}$.
The sum of the bound masses of a group in a supercluster gives the mass of a supercluster. 
Therefore, in these calculations, the mass
of groups takes into account both masses of galaxies and masses of dark matter haloes.
These masses may still be underestimated as
the mass in very poor groups and single galaxies, and the gas mass is not taken into account.

The masses of superclusters  in \citep{2024A&A...683A.130L}, based on eROSITA clusters,
are determined as a sum of virial masses of clusters,  $M_{\rm 200}$, which are calculated
using masses $M_{500}$ of its member clusters, 
 $M_{\rm 200}\approx 1.46\times M_{500}$
as described in detail in \citep{2024A&A...683A.130L}. To calculate 
total supercluster masses, the authors use a relation based on simulations by \citep{2014A&A...567A.144C},
$M_{\rm cl} = 0.39 \times M_{scl}$, where 
$M_{cl}$ is the sum of virial masses ($M_{\rm 200}$) of clusters within a supercluster.

The masses of groups have been used  to calculate the supercluster 
masses in several studies \citep{2015A&A...580A..69E, 2016A&A...595A..70E, 2022A&A...668A..37H, 
2024ApJ...975..200C}.
When calculating the mass of superclusters using group and cluster masses, one source of
uncertainty is related to how precisely the masses of groups and clusters have been found.
Ref. \citep{2015A&A...580A..69E} used several mass estimations for galaxy groups. Namely, they used
virial masses of groups calculated by \citep{2014A&A...566A...1T} ($M_{\rm 200}$), and for poor groups for which 
virial masses are poorly defined, they used the relation between the stellar mass of the brightest group
galaxies and the halo mass \citep{2010ApJ...710..903M}. 
Another possibility is to use for poor groups the median mass
of these groups, multiplied by the number of these groups in a supercluster. Ref. \citep{2015A&A...580A..69E}
compared these two methods and found a good agreement in mass estimation.
The mass of single galaxies has been taken into account using several different 
approaches. Single galaxies may be the brightest galaxies of faint
groups, for which other group members are too faint to be included in flux-limited redshift
surveys. The mass of such groups can be calculated using the galaxy mass--halo mass relation
\citep{2010ApJ...710..903M} or by using the median mass of very poor groups as a proxy of their
mass. These mass estimated agree well \citep{2015A&A...580A..69E, 2016A&A...595A..70E}. 
 
Approximately 10\% of supercluster mass comes from intercluster
gas \citep{2012Natur.487..202D, 2016A&A...592A...6P}. 
Therefore, supercluster masses can be calculated 
by summing the masses of groups and clusters, and adding the estimated gas mass in superclusters
\citep{2016A&A...592A...6P, 2016A&A...595A..70E, 2021A&A...649A..51E}. 
In total, ref. \citep{2021A&A...649A..51E} estimated that the contribution of
rich groups and clusters to the total mass of superclusters is  approximately 75\%
of the total mass, and 10\% and 5\% of the mass comes from poor groups and single galaxies,
correspondingly. 
The remaining mass comes from the gas in the supercluster.
These studies  have shown that while the masses of poor superclusters
may be of the order of a rich galaxy cluster, $M \approx 1\times~10^{14}M_\odot$,
the masses of the richest superclusters may be of the order of
$M \approx 1\times~10^{16}M_\odot$ or even higher.

From observations, one can determine the luminosity of groups and clusters in a supercluster.
Also, for superclusters determined using the luminosity-density field, the total luminosity 
of superclusters can be calculated. Therefore, the mass of a supercluster can be calculated using 
the luminosity of a supercluster, and the mass-to-light ratio $M/L$.
In this way the mass in dark matter is also taken into account. However, at first, we should know
the characteristic mass-to-light ratio for superclusters, its possible dependence on supercluster 
richness, and perhaps on other properties of superclusters. These problems were discussed in
\citep{2015A&A...580A..69E, 2016A&A...595A..70E}.  Ref. \citep{2014MNRAS.439.2505B} showed that the mass-to-light ratio
of galaxy groups is $M/L \approx 400$, and this ratio is higher for poorer groups.
They also found that $M/L$ of groups does not depend on the global environment of
groups. However, ref. \citep{2015A&A...580A..69E} showed that  $M/L$ of groups is higher in supercluster
outskirt regions. They attributed this to the fact that these regions are mostly populated 
by poor groups while rich clusters with lower $M/L$ values preferentially lie in
the high-density cores of superclusters. Therefore, while poor superclusters have 
mass-to-light ratio $M/L \approx 400$, rich superclusters have lower $M/L$ values,
$M/L \approx$ 250--300 \citep{2015A&A...580A..69E, 2016A&A...595A..70E, 2022A&A...668A..37H}. 
Ref. \citep{2022A&A...668A..37H} speculated that perhaps poor superclusters embed a 
higher fraction of dark matter than rich superclusters.
Comparison of the masses of superclusters calculated from the masses of member groups,
and using the total luminosity of groups and these $M/L$ values showed a good agreement 
\citep{2015A&A...580A..69E, 2016A&A...595A..70E, 2017A&A...603A...5E}.

As described in Section~\ref{sect:morph}, using PCA, one can derive the scaling relations
for superclusters that combine morphological and physical parameters of superclusters
to obtain supercluster luminosities. This provides an independent method to calculate
supercluster masses using the luminosity found with scaling relations, and $M/L$ ratios
obtained in previous studies. Ref. \citep{2017A&A...603A...5E} applied this method to estimate
the masses of the BOSS Great Wall superclusters. Masses, calculated in this way, agreed well
with the masses of these superclusters found  using luminosities and stellar 
masses of galaxies.

Supercluster masses have also been estimated using data on supercluster volumes
and their overdensity. 
In this method, the mass of superclusters is found using data on the 
volume of a supercluster, and its overdensity compared to the mean
cosmic density~\citep{2016A&A...588L...4L, 2021A&A...656A.144B}.
Under the assumption that the mass 
density correlates with luminosity density,
ref. \citep{2016A&A...588L...4L} divided the volume of superclusters into
cells, and calculated the mass of superclusters as the sum of the mass 
of each grid cell as $M=\rho_c D V$, 
where $D$ is the luminosity density in a cell in the units of the mean densities,  
$V=$ is the volume of the cell, and $\rho_c$ is the critical density of the Universe.

One source of the difference in supercluster masses from different studies 
can be attributed to the difference in the supercluster definition. 
For example, while \citep{2024ApJ...975..200C} define superclusters as future collapsing
systems, ref. \citep{2023ApJ...958...62S} define superclusters as  relatively isolated
high-density regions (superclusters with high-density cores surrounded by lower-density outskirts
regions). As a consequence,
the highest-mass superclusters in these studies have masses of the order of
$M \approx 1\times 10^{15}~h^{-1}M_\odot$ in \citep{2024ApJ...975..200C}
and $M \approx 2.5\times 10^{16}M_\odot$ in~\citep{2023ApJ...958...62S}.

%%%%%

\section{Superclusters as Biased Traces of the Mass in the Cosmic Web}
\label{sect:bias}

Biases in mass estimations related to the differences between observed mass and total mass 
are related to the more general problem of how well the visible matter traces the distribution of dark matter
in the Universe. The problem of the relative distribution
of visible and dark matter is usually referred to as a biasing problem, introduced by 
\citep{1984ApJ...284L...9K}. Namely,
galaxy clusters, especially clusters in superclusters 
represent high-density peaks in the cosmic density field, thus being 
biased tracers of this field \citep{1996astro.ph.11148B, 2024MNRAS.529.4325B}. 
However, the emphasis in studies of the biasing problem is mostly on the study of the
distribution of matter in low-density regions of the cosmic web---voids \citep{2001ApJ...557..495P}. 
For example, using $N$-body simulations, the hierarchical nature of the
cosmic web was discussed by \citep{2024MNRAS.527.4087J}. The authors defined cosmic web
elements as nodes (galaxy groups and clusters), filaments, walls, and voids,
and studied the properties of galaxies in various environments. 
This study showed, among other results, that in high-density environments
(in our context---in superclusters, although this was not specifically addressed 
in this study)
the distribution of low-mass haloes follows the distribution of high-mass haloes. 

Next, I focus on the relative distribution of visible and 
dark matter in superclusters.
First of all, the total mass of superclusters  can be estimated using simulations. 
For example, if the mass of
superclusters is determined using masses of galaxy groups and clusters embedded in
superclusters, then the total mass of superclusters, obtained in this way, 
may be underestimated if the mass in poor groups and
single galaxies, as well as the mass of gas in superclusters is not taken into account
\citep{2014A&A...567A.144C}. They estimated that depending on the mass (richness) limit
of groups and clusters 
used to determine superclusters, the bias factor introduced in mass estimation may be of the order of
1.6--1.8. 

The comparison of the mass of SCl~A2142 based on group
masses with these predictions from simulations
showed that the bias value was slightly lower---1.5 \citep{2015A&A...580A..69E}. 
In this study, the mass in poor groups and
single galaxies, as well as the mass of gas in the supercluster, were taken into account, and this may be the reason 
for the lower bias value. 

Review paper \citep{1996astro.ph.11148B} provides extensive discussion on the masses
and mass-to-light ratios of galaxy clusters. In particular, the author addresses the 
question of possible extra dark matter in clusters, in addition to the dark matter related 
to dark matter haloes of galaxies in clusters. Based on masses and luminosity of clusters,
it is shown that typical mass-to-light ratios of galaxy clusters are $M/L \approx$ 200--300,
and there is no additional dark matter. The same conclusion is made for superclusters.

Observational proof of the presence of possible additional dark matter in superclusters
may come from the studies of supercluster masses from lensing surveys.
One example is the analysis of the distribution of 
mass and light in the Abell 901/902 supercluster at redshift $z = 0.165$ \citep{2008MNRAS.385.1431H}.
This  supercluster embeds  two galaxy clusters connected by a filament of galaxy groups.
The distribution of dark matter in the A901/902 supercluster was analysed using a weak lensing analysis.
This analysis showed that  visible galaxies and clusters
in the supercluster trace well the distribution of
dark matter, with no need for extra dark matter.
 
Also, in the study of the King Ghidorah supercluster, the question of relative  
distribution of stellar mass and
dark matter was addressed  \citep{2023MNRAS.519L..45S}.
The distribution of dark matter was determined using weak lensing signals. The distribution 
of stellar mass was found using the stellar mass in massive galaxies in the supercluster.
The comparison of these two maps revealed strong correlation between these two.
The authors concluded that visible mass follows the distribution of dark matter.
This conclusion was supported by the analysis of mock catalogues.

These examples lead to the same conclusion: up until now, there is no strong evidence
of the presence of extra dark matter in superclusters, which may strongly bias the estimation of supercluster
masses.

%%%%%%%%%%%%%%%%%%%%%%%%%%%%%%%%%%%%%%%%%%%%%%%%%%

\section{Supercluster Structure---High-Density Cores and Outskirts}
\label{sect:cores}

When analysing the density distribution in superclusters of various richness, determined using the 
luminosity-density field of galaxies, refs. \citep{2007A&A...464..815E, 2007A&A...462..397E} 
noticed important differences 
between rich and poor superclusters. While rich superclusters had high-density central parts
or cores, in poor superclusters, even maximal densities were lower, typical of the outskirt regions 
of rich superclusters. The division between rich
and poor superclusters was approximately the same, as later found using PCA, as described in 
Section~\ref{sect:morph}. At the border of high-density cores (hereafter HDCs), the morphology 
of superclusters changes~\citep{e07, 2011A&A...532A...5E}.
The analysis of the structure of superclusters in the Sloan Great Wall with normal
mixture modelling also showed that these superclusters have several high-density cores~\citep{2016A&A...595A..70E}. 
Ref. \citep{2024PASA...41...78Z} called such regions nucleation regions in superclusters,
and they presented a list of such regions in nearby superclusters. 
Cores in their study were defined
as large gravitationally bound regions in superclusters, 
which may host two or more clusters and groups, and where the density of matter is high enough to
survive cosmic expansion and collapse in the future.

Typically, HDCs of superclusters embed one or several rich clusters,
connected by filaments of poor groups and galaxies. The overall masses of 
HDCs are of the order of \mbox{$M \approx$ 0.5--$5\times 10^{15}M_\odot$,} i.e., of the
same order as the masses of superclusters determined using the future collapse criterion
\citep{2022A&A...666A..52E, 2024PASA...41...78Z}. Ref. \citep{2022A&A...668A..37H} noted that the most spherical superclusters
from the adaptive supercluster catalogue by \citep{2012A&A...539A..80L} correspond to the
high-density cores of superclusters from the catalogue, in which superclusters were defined using
fixed density threshold.

The detailed analysis of the mass distribution in the HDCs of superclusters
reveals a small density minimum around the central
clusters of HDCs. Refs. \citep{2020A&A...641A.172E, 2021A&A...649A..51E} showed that this minimum
marks the borders of the region of influence around clusters.
Within the regions of influence, all galaxies, groups, and filaments are
falling into clusters.  The radius of the spheres of influence has been called
depletion radius in \citep{2021MNRAS.503.4250F}.
The density contrast at the borders of the regions of influence is of the order $\Delta\rho_{inf}$ = 30--40.
Using the spherical collapse model, ref. \citep{2021A&A...649A..51E} showed
that the regions of influence were at turnaround and started to collapse at redshifts $z \approx$ 0.3--0.4.

Another characteristic density contrast in the matter distribution
around clusters is the turnaround density contrast. In the $\Lambda$CDM
Universe with non-zero cosmological constant  $\Delta\rho_{turn} = 13.1$
\citep{2015A&A...581A.135G, 2015A&A...575L..14C} (and references therein).
Typical radii of turnaround regions in the HDCs of superclusters are 
of the order of $R_{turn} \approx$ 5--10 \Mpc. However, in the richest superclusters
in the nearby Universe, the Shapley and the Corona Borealis superclusters,
the radius of the turnaround region is even larger, $R_{turn} \approx$ 12--13~\Mpc.

\section{Fractal Properties of Superclusters}
\label{sect:frac}

The cosmic web can be described as a multifractal pattern,
with  different  fractal dimensions at different scales
\citep{1982fgn..book.....M, 1997cdc..conf...24P, 
1998BrJPh..28..132R, 2002sgd..book.....M, 2005astro.ph..5185B}.
One method to find the fractal dimension of the cosmic web is provided 
by the correlation analysis \citep{2002sgd..book.....M}. 
Many studies have shown that the slope of the correlation function
changes at a certain scale, which shows the crossover from correlations
in galaxy groups and clusters to the correlations between galaxies in
the filamentary pattern connecting clusters; in other words, from
one fractal regime to another, with a different fractal dimension
which also means a crossover from a non-linear dynamical regime at small scales 
to linear dynamical regime at large \mbox{scales~\citep{1996astro.ph.11148B, 2002sgd..book.....M, 2005astro.ph..5185B, 2020A&A...640A..47E}.}
Fractal dimensions can be calculated from the slope of the correlation
function $\xi(r)$, which give us the so-called structure function,
$g(r)=1 +\xi(r)$, and the fractal dimension \mbox{$D(r)= 3+ d \log g(r)/ d \log r$,} 
where $r$ denotes distance between \mbox{galaxy pairs}.

At scales up to $r \approx$ 3--10~\Mpc\ (depending on the sample under
study) \mbox{$D \approx$ 1--1.5,} telling that the structures are more one-dimensional.
At larger scales $D \approx$ 2--3, showing the crossover to more two-dimensional structures
\citep{2002sgd..book.....M, 2020A&A...640A..47E}.
Ref. \citep{1997A&AS..123..119E} showed that fractal dimension of clusters in
very rich superclusters has a value $D \approx 1.4$, and fractal dimension
determined using all clusters is $D \approx 2$. 
Among individual superclusters, the 
fractal properties of the Saraswati supercluster were analysed by \citep{2019InJPh..93.1385R}.
This study applied the so-called box-counting method to estimate the 
fractal dimension of the supercluster. They found that $D \approx$ 1.7--2,
showing that the Saraswati supercluster is essentially a \mbox{planar structure.}

\begin{figure}[H]\vspace{-9pt}
\centering
%\isPreprints{\centering}{} % Only used for preprints
\begin{adjustwidth}{-\extralength}{0cm}
\subfloat[\centering]{\includegraphics[width=7.7cm]{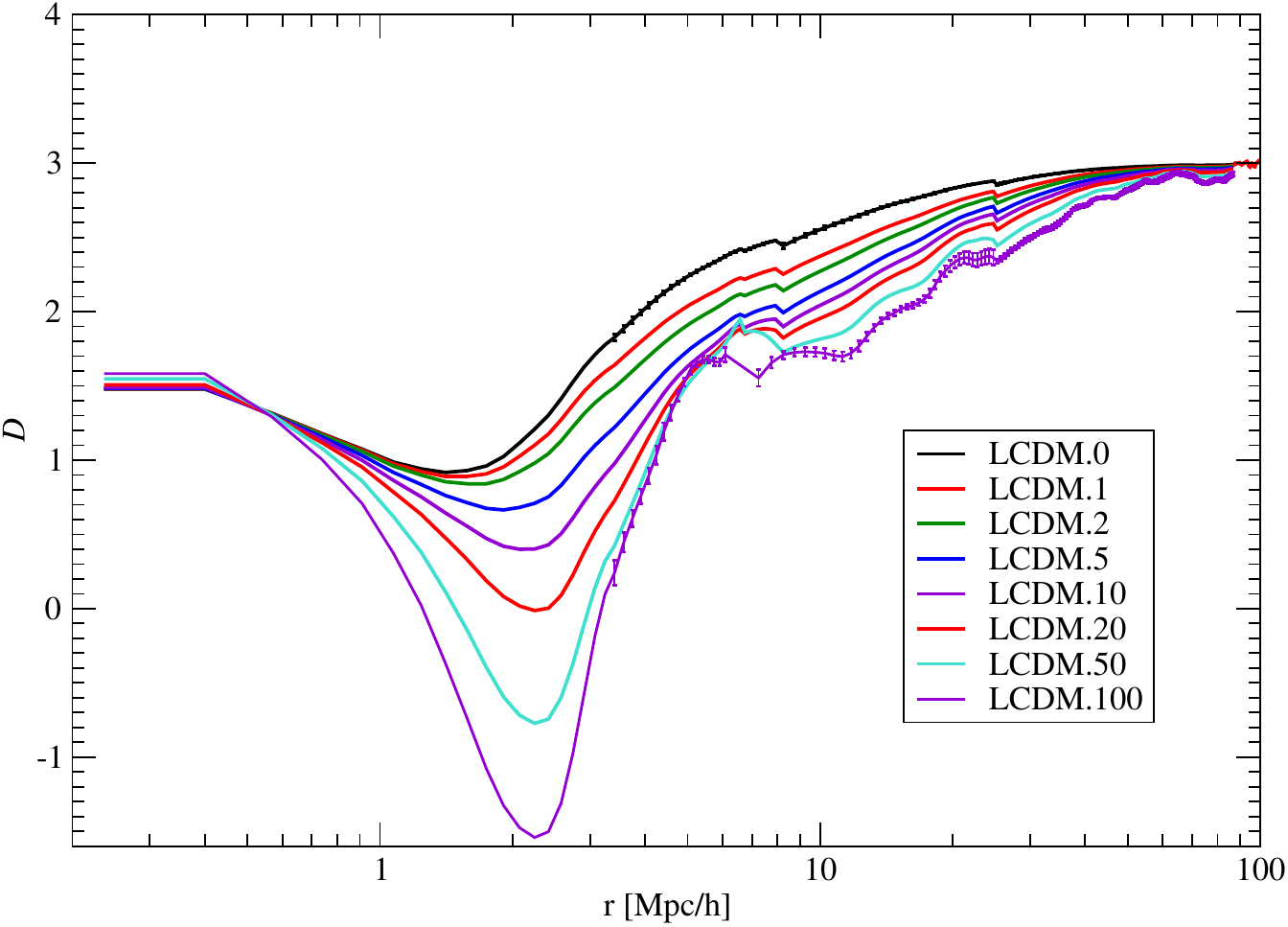}}
\hfill
\subfloat[\centering]{\includegraphics[width=7.7cm]{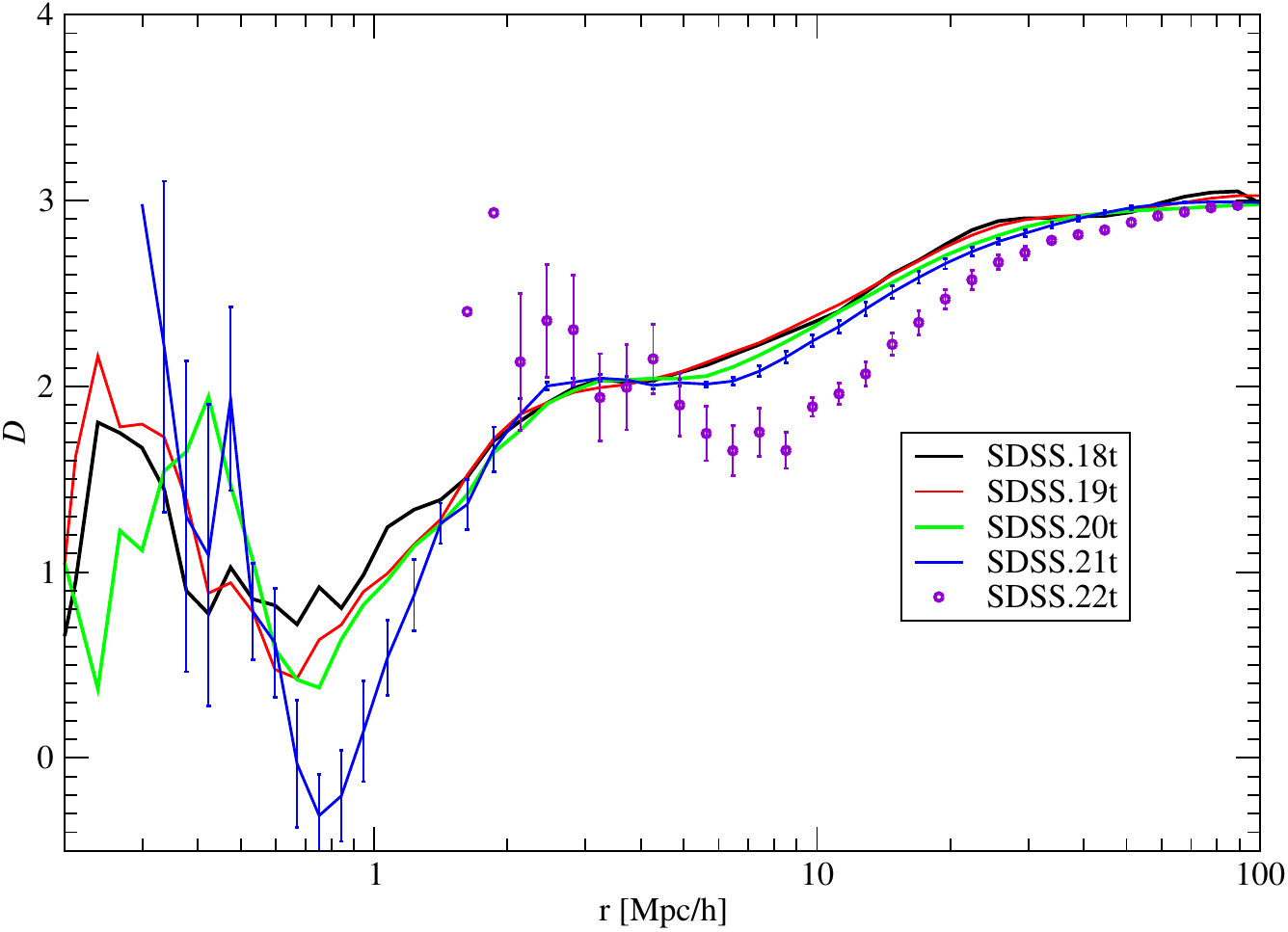}}\\
\end{adjustwidth}
\caption{
Fractal dimension functions, $D(r)=3+\gamma(r) $
for $\Lambda$$CDM$ models for different particle selection limits (a), 
and for
SDSS galaxies with five luminosity thresholds (b).
Figure by Jaan Einasto~\citep{2020A&A...640A..47E}.  
}
\label{dfrac}
\end{figure}

The fractal dimension at different scales is shown in Figure~\ref{dfrac},
which presents the fractal dimensions on the left panel, calculated using particle data in
the $\Lambda$$CDM$ model, calculated in a box with size $512$~\Mpc\ with various particle density
limits, as shown in the figure. The right panel of Figure~\ref{dfrac} shows the fractal dimension of
SDSS galaxy samples with various luminosity limits.

At separations $r \le 3$~\Mpc,  correlation and fractal functions of galaxies 
reflect the galaxy or dark matter distribution in groups and clusters within halos. 
The fractal dimension changes within limits $-1 \le D \le 1.5$. 
At larger distances, the fractal dimension gradually  increases from 
1 to 3 at the largest distance studied, 100~\Mpc.  
Therefore, the fractal dimension is not constant, as in simple fractal models.

In the
small separation region, the spatial correlation function of the LCDM model
samples is proportional to the density of matter, and measures the
density profile inside the DM halos.  
The fractal dimension function of the LCDM
samples depends on the particle density limit $\rho_0$, but still, the crossover from
one fractal regime to another occurs approximately at distances $r$ = 5--6~\Mpc.
The fractal dimension of the SDSS sample behaves similarly, but note that the change in fractal dimension
of the brightest galaxies with $M < -22$ occurs at larger distances than in the case of lower
luminosity galaxies. The brightest galaxies in the sample are mostly the brightest galaxies in clusters
in superclusters, and the change in the fractal dimension is related to the crossover from clusters 
in the supercluster core regions to outskirts. 
As shown above, at high-density cores of rich superclusters, the 
morphological signature of superclusters also changes. 
The possible connection of these two changes may be one direction in the
future studies of superclusters.

At still larger scales, up to over 100~\Mpc\, the correlation function of all galaxies 
is featureless, except, of course, the famous Baryon Acoustic Oscillation (BAO) feature
\citep{2005ApJ...633..560E}. Therefore, at large scales, 
even in samples with very large structures, the fractal dimension does not
change \citep{2009EL.....8649001S}. 
This is often interpreted as a signature of the homogeneity of the galaxy
distribution at  scales larger than this limit, 100~\Mpc. However, this interpretation
has been questioned \citep{1997ApL&C..36...65P}, and it 
is not in agreement with the presence of very large  underdense regions (voids) and superclusters, 
with sizes over 100~\Mpc, as, for example, superclusters in the Sloan Great Wall and in the BOSS 
Great Wall, the Bootes void, the Eridanus supervoid, associated with the CMB Cold Spot,
and other voids \citep{1981ApJ...248L..57K, 2012ApJ...759L...7P, 2012A&A...539A..80L, 
2015MNRAS.450..288S, 2016MNRAS.462.1882K, 2022MNRAS.515.4417K}.
Very large voids detected in the distribution of luminous galaxies  
or galaxy groups/clusters are not empty, but filled with the hierarchical, filamentary
web of fainter galaxies, in which the sizes of smaller voids 
between these structures depend on the luminosity limit of galaxies
used to determine these 
voids \citep{1987ApJ...314..493K, 1995A&A...301..329L, 1996A&A...314....1L, 
2023A&A...673A..38C, 2024MNRAS.527.4087J}.

The correlation function may be similar for samples of very
different geometry, as it does not contain phase information, as shown in detail in \citep{2009LNP...665..493C}.
As discussed \mbox{in \citep{1997ApL&C..36...65P, 2005astro.ph..5185B, 2009EL.....8649001S, 2020A&A...640A..47E},} 
the two-point correlation function is 
self-averaging up to approximately $30$~\Mpc, and at these scales and higher, it is characterised by a fractal
dimension $D \approx$ 2--3, and this cannot be interpreted as a crossover to homogeneity
\citep{1997ApL&C..36...65P}.
This conclusion agrees with observational evidence on the lack of
homogeneity in the distribution of superclusters and voids 
at scales at least up to 300~\Mpc\ 
 in the nearby \mbox{universe \citep{2009EL.....8649001S, 2012ApJ...759L...7P,
2015JKAS...48...75P, 2025arXiv250201308C, 2025arXiv250401669D}}.

As mentioned, underdense regions (voids) in the cosmic web are of a hierarchical nature.
Voids determined using luminous galaxies may be divided into smaller voids by filaments
of fainter galaxies, but this does not change the fractal dimension of the samples under
study. For example, in Figure~\ref{dfrac} (right panel), the fractal dimension
of SDSS galaxy samples is the same for galaxy samples
with  a wide range of magnitude limits, \mbox{$M$ = $-$18--$-$21}, while the mean diameters
of voids delineated by galaxies with $M < -20$ are almost twice as large
as mean diameters of voids delineated by fainter galaxies, $M < -18.8$,
23 versus \mbox{13~\Mpc\ \citep{1995A&A...301..329L}} (see also \citep{2024MNRAS.527.4087J}).
These void size values agree well with the determinations of void sizes based
on SDSS data and galaxy magnitude limit $M_r < -20$~\citep{2023ApJS..265....7D}, who found that
median effective void sizes are in the range of 15--19~\Mpc, based on different
void finding algorithms.
The sizes of voids depend also on the sample size~\citep{2014MNRAS.440.1248N}.

\section{Supercluster Evolution in the Cosmic Web}
\label{sect:evol}

The formation and evolution of the cosmic web 
under the gravity of dark matter
and the acceleration from dark energy can only be followed using simulations 
\citep{2014MNRAS.441.2923C}.
Simulations show that 
the formation of the present-day structures in the cosmic web began with the growth of
tiny density perturbations in the very early Universe under gravity
\citep{1978MNRAS.183..341W, 1980Natur.283...47E, 1996Natur.380..603B}. 
During evolution, positive dark matter density perturbations grow by merging and infall 
of surrounding proto-structures, and as a result, the present-day cosmic web consists
of structures of various properties and characteristic scales, from dwarf galaxies to
rich clusters, filaments, and voids between them.  
Superclusters, which host rich galaxy clusters, connected by filaments of
poor groups and single galaxies,  form where positive sections
of medium- and large-scale dark matter  density perturbations combine. 
Dark matter perturbations of a scale of about 100 \Mpc\ give rise to the largest superclusters.
An analysis of the simulations of the future evolution of the cosmic web and various 
structures in it in the $\Lambda$$CDM$
universe with a non-zero cosmological constant shows that in the distant future,
structures in the cosmic web will be of higher contrast, superclusters will become more spherical 
and compact, and voids will become emptier \citep{2007JCAP...10..016H, 2009MNRAS.399...97A}.

\begin{figure}[H]
%\isPreprints{\centering}{} % Only used for preprints
\includegraphics[width=14 cm]{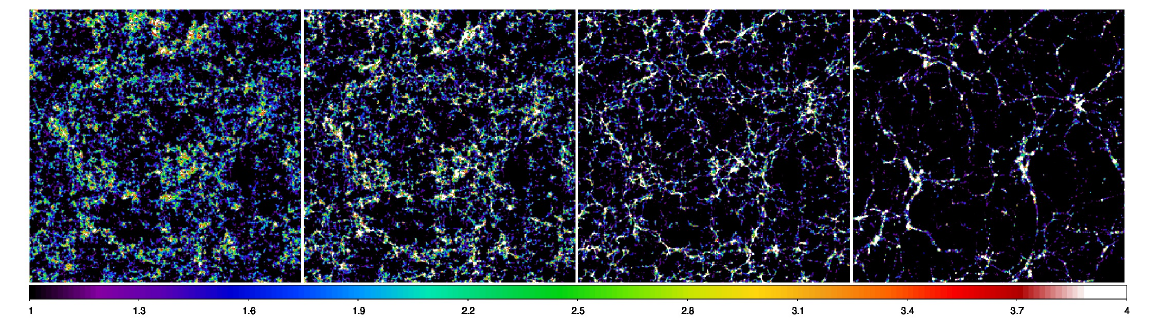}
\includegraphics[width=14 cm]{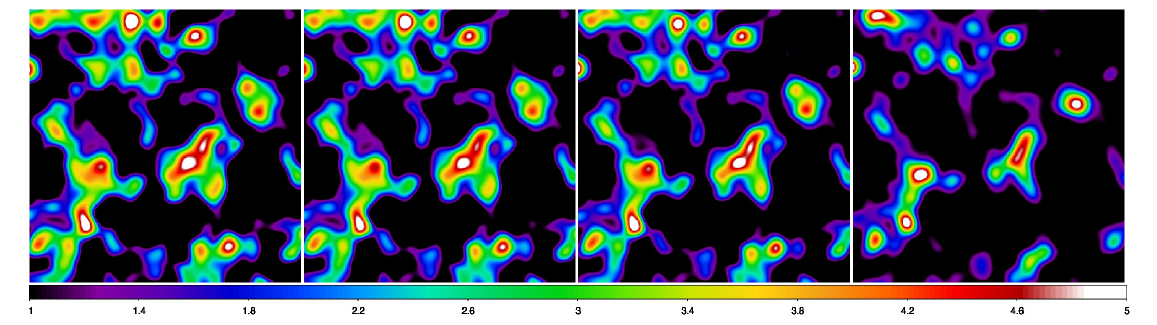}
\caption{
Density fields of simulation in the $256$~\Mpc\ box without additional smoothing (upper panels) and with
  a smoothing length  $8$~\Mpc\ (lower panels). The panels
  from left to right show fields for epochs $z=30$, $z=10$,  $z=3$, and
  $z=0$,  presented in  slices of size   $200\times200\times0.5$~\Mpc. 
  Only overdensity regions are shown
  with colour scales from left to right  1--1.4, 1--2, 1--4, 1--8 in the (upper
  panels) and 1--1.08, 1--1.25, 1--1.8, 1--5 in the (lower panels).
Figure by Jaan Einasto \citep{2021A&A...652A..94E}.  
\label{denevol}}
\end{figure}

The whole cosmic web can be divided between
the  BoAs (superclusters with cocoons)   (see Section~\ref{sect:def}). 
Superclusters fill only a small 
fraction of BoA space, about 1\% at the present epoch.
To understand the evolution of various populations in the cosmic web, their evolution
should be analysed simultaneously. Various populations
in the dark matter density field  can be defined, using different smoothing
lengths. 
The evolution of cluster-size and supercluster-size populations in the density field 
in a redshift interval \mbox{$z$ = 30--0} was followed by \citep{2019A&A...623A..97E}.
Figure~\ref{denevol} shows four snapshots from the simulations,
which were performed in the box size $256$~\Mpc\ with $512^3$ particles
using the {\sc GADGET} code. Figure~\ref{denevol}  presents snapshots
at redshifts $z=30$, $z=10$,  $z=3,$ and $z=0$, in the upper panel
without smoothing, and in the lower panel, with a smoothing length  8~\Mpc,
which has also been used to determine observed superclusters using the density field
method (Section~\ref{sect:def}).
High-density regions in the lower panel represent superclusters, and low-density
regions represent voids or supercluster cocoons between superclusters.
%EE: Please check that the intended meaning has been retained  %ME Made minor changes for clarity
 The figure shows that the 
supercluster locations and the  BoA  borders do not change much during 
the evolution of the cosmic 
web, and an essential evolution of various populations  in BoAs  occurs 
at smaller scales within superclusters and their cocoons.
%
%%%%%%%%%%%%%%%%%%%%%
Volumes in comoving 
coordinates and masses of BoAs remain approximately constant during evolution, 
whereas masses of superclusters increase during evolution by a factor 
of about 3 by the infall of surrounding matter inside BoAs to superclusters. 
The most massive $\Lambda$CDM superclusters have, at the present epoch, masses 
$M_{scl} \approx 10^{16}M\odot${, and the most massive supercluster basins have at all epochs masses}
$M_{basin} \approx 2\times 10^{17}M_\odot$. 
The exchange of matter 
between neighbouring BoAs is minimal, 
because the velocity flows within BoAs is directed inwards.
%%%%%%%%%%%%%%%%%%%%%%%%%%%%%%%%%%%5

Using observational data, the future evolution of  individual high-density cores of 
superclusters, and/or the evolution
 of full superclusters, as well as the dynamical state of superclusters,  are most commonly  analysed using
the spherical collapse model \citep{1980lssu.book.....P}.
This model follows the evolution of a 
spherical shells, determined by the dark matter mass
in its interior. The evolution of spherical shells 
 can be characterised by several important
epochs. To analyse the evolution of superclusters, important
epochs are the turnaround and the future collapse \citep{2015A&A...575L..14C}.
The  spherical overdensity within  a radius $R$ can be described as $\Delta\rho \equiv \rho / \rho_m$,
where $\rho$ is the matter density in the volume, and $\rho_m$ is the mean matter density. 
This overdensity expands together with the Universe, but because of  its internal gravitation, 
it expands more slowly than the surrounding background.

At a certain epoch, this (slower) expansion may stop, and the region starts to contract. 
This epoch is called a turnaround. 
Depending on the value of the density contrast $\Delta\rho$, 
the turnaround may occur in the past, at present, or in the future, or may not occur at all. 
In an expanding $\Lambda$$CDM$ universe, the turnaround density contrast
at the present (at redshift $z = 0$) is $\Delta\rho_{turn} = 13.1$.
If a spherical density perturbation is not high enough to collapse at present,
it may still collapse in the future, if its density contrast (future collapse density contrast)
now (at redshift $z = 0$) is $\Delta\rho_{FC} = 8.73$. 
The density contrast $\Delta\rho_{ZG} = 5.41$ corresponds to so-called zero 
gravity (ZG), 
at which the radial peculiar velocity component of 
test particle velocity equals the Hubble expansion
and the gravitational attraction of the system and its expansion are equal.
The density contrast 
$\Delta\rho = \rho/\rho_{\mathrm{m}} = 1$ 
corresponds to the linear mass scale or the Einstein--Straus radius at which
the radial velocity around a system reaches the Hubble velocity,
$u = HR$ and peculiar velocities $v_{\mathrm{pec}} = 0$
\citep{2015A&A...577A.144T, 2015A&A...581A.135G}. 
This scale approximately corresponds to the cocoon boundaries
\citep{2015A&A...575L..14C, 2015A&A...577A.144T, 2015A&A...581A.135G}.
The mass $M$ of a density perturbation within a sphere with radius $R$ can be calculated as
\begin{linenomath}
    \begin{equation} 
(R)=1.45\cdot10^{14}\,\Omega_\mathrm{m0}\Delta\rho\left(R/5h^{-1}\mathrm{Mpc}\right)^3 \cdot (1+z)^3 h^{-1}M_\odot,
\label{eq:masrad}
    \end{equation}
\end{linenomath}
where $\Omega_\mathrm{m0}$ is the matter density and $\Delta\rho$ denotes density contrast.

The density contrast at turnaround depends on cosmological parameters. For example,
for $\Omega_{\mathrm{m}} = 1$, 
the density perturbation at the turnaround is 
$\Delta\rho_{\mathrm{T}} = (3\pi/4)^{2} = 5.55$ 
(the density contrast $\delta_{\mathrm{T}} = \Delta\rho_{\mathrm{T}} - 1 = 4.55)$
\citep{2002sgd..book.....M}.

Superclusters, which have been defined
as future collapsing systems, will collapse by definition. 
In superclusters with high-density cores surrounded by lower-density outskirts
regions, the high-density cores  may reach turnaround at present, and
start to collapse during future evolution, if the density contrast in their cores
is as high as $\Delta\rho \geq \Delta\rho_{FC}$.

To apply the spherical collapse model, at first, one must calculate the distribution of mass 
in the supercluster, centred at its centre of mass (most massive cluster; in superclusters with 
several high-density cores, this distribution is calculated for each core).
Based on the mass distribution, one can find the density contrast around the central 
cluster, and this can be directly compared with the characteristic density contrasts
from the spherical collapse model. This approach has been used to analyse the evolution of
superclusters, as the Shapley, the Laniakea, the Corona Borealis,
the Sloan Great Wall, the BOSS Great Wall, the Perseus-Pisces,
 the Hercules, the Coma, the Saraswati, and \mbox{others (\citep{2000AJ....120..523R,
2006MNRAS.366..803D, 2014MNRAS.441.1601P, 2015A&A...575L..14C,
2015A&A...580A..69E, 2015A&A...577A.144T,2015MNRAS.453..868O, 2016A&A...595A..70E, 2017ApJ...844...25B, 2018A&A...619A..49C, 
2021A&A...649A..51E, 2024PASA...41...78Z}} and references therein).
Ref. \citep{2024PASA...41...78Z} provides a list of 105 cores in 53 rich superclusters in the nearby
Universe. They emphasise that these cores are the most massive structures that may 
be collapsing now or in the future.

\begin{figure}[H]
%\isPreprints{\centering}{} % Only used for preprints
\includegraphics[width=10 cm]{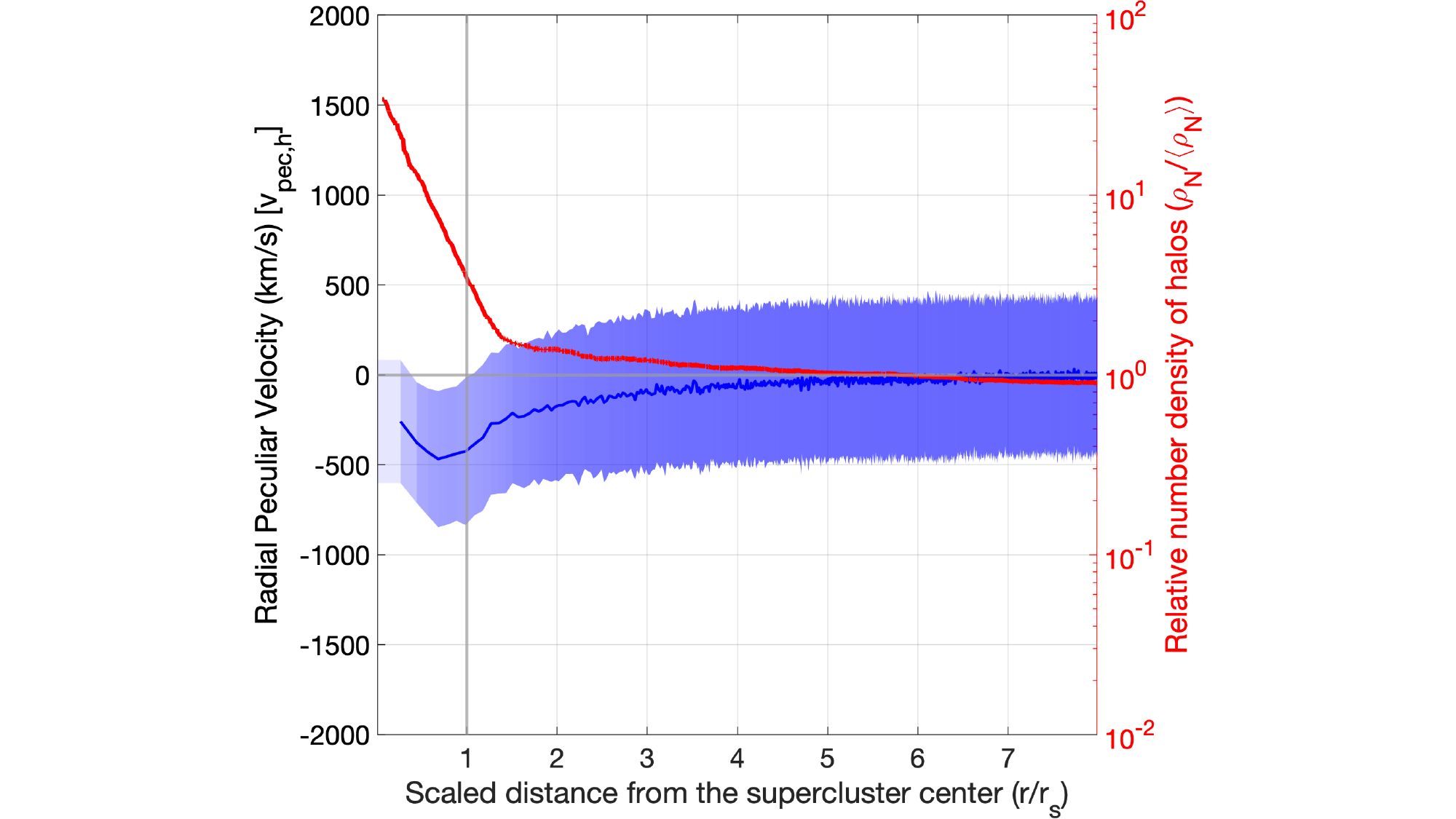}
\caption{Mean %MDPI: Please change the hyphen (-) into a minus sign (−, “U+2212”) in the figure, e.g., “-1” should be “−1”.
  values of the radial peculiar velocity component of haloes (blue line)
	as a function of the  scaled distance  from the supercluster center. 
	The shaded blue region shows the 1~$\sigma$
	standard deviation of the radial peculiar velocity component of haloes. 
	The red line shows the relative number density of halos as a function of scaled radius, 
	calculated by dividing the average number density $\rho_N$ (of all stacked superclusters) at a scaled radius 
	by the total number density $\langle\rho_N\rangle$ of haloes within a 
	sphere of radius $\sim 8 \times r_s$. The x-axis denotes the scaled distance from the supercluster center in units of
	$r_s$ (half the size of a supercluster). Figure by Shishir Sankhyayan \citep{2023ApJ...958...62S}.  
\label{phase}}
\end{figure}

The detailed analysis of the dynamics of groups and clusters in superclusters
and in their high-density cores requires data on
the peculiar velocities of galaxy groups and clusters (dark matter haloes) in superclusters. These data are available for 
simulated superclusters. Therefore, Figure~\ref{phase} shows radial peculiar velocities of dark matter haloes
in superclusters versus the scaled distance from the supercluster centre 
(in units of $r_s$, half the size of a supercluster) (phase space diagram)  
by \citep{2023ApJ...958...62S}. In this figure, more than 90\% of supercluster member halos 
have a negative supercluster-centric peculiar velocity component. This means that their movements are
influenced by the gravitational potential of 
their host supercluster. This influence is weak in supercluster outskirts, at
large scaled distances, and negligible on the halos  outside the supercluster region.
The velocities of dark matter haloes at scaled distances half the size of superclusters (the core regions)
are directed toward supercluster centres, and this may lead to the collapse of these centres in the future.

\begin{figure}[H]
%\isPreprints{\centering}{} % Only used for preprints
\includegraphics[width=10 cm]{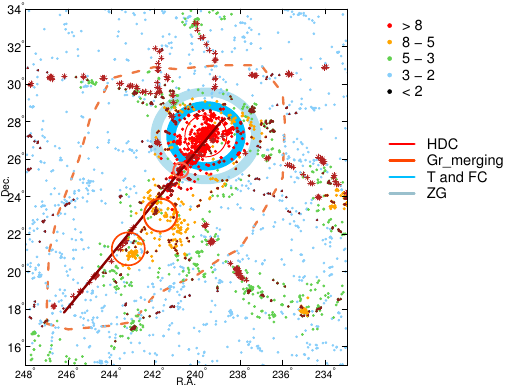}
\caption{Sky  distribution of galaxies in and around 
the supercluster SCl~A2142. 
Coloured dots  denote galaxies in regions
of different luminosity-density $D8$ as indicated in the legend. 
Galaxies in long filaments with length $ \geq $20~\Mpc\
are denoted with stars of dark red colour. HDC of the supercluster is marked with a dark red circle.
Orange circles mark the location of merging groups, which will separate from the supercluster in the future.
Blue circle marks turnaround (T) region, 
and light blue circle shows borders of zero gravity (ZG). The future collapse (FC) region 
lies between these regions. The dark red line denotes the supercluster axis.
The orange dashed line shows the cocoon boundaries, 
determined using minima in the density field to study
the cocoon region of the supercluster SCl~A2142. Figure by ME \citep{2020A&A...641A.172E}.
\label{a2142sky}}
\end{figure}

\begin{figure}[H]
%\isPreprints{\centering}{} % Only used for preprints
\includegraphics[width=8 cm]{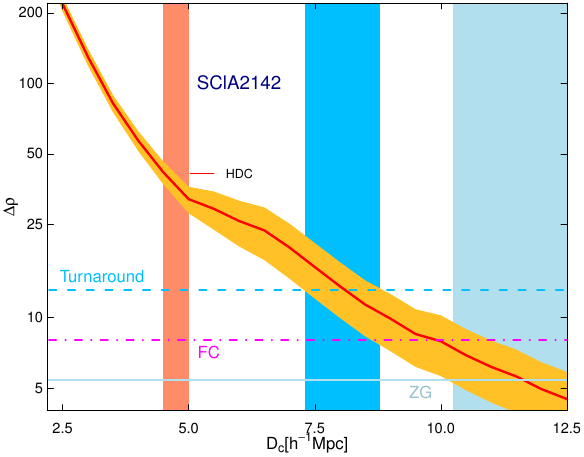}
\caption{Density contrast $\Delta\rho = \rho/\rho_{\mathrm{m}}$ versus
supercluster-centric distance $D_c$ for the SCl~A2142 main body (red line). 
Golden area shows error corridor from mass errors. Characteristic density contrasts 
are denoted as follows: $\Delta\rho = 13.1$ (turnaround, blue dashed line),
$\Delta\rho = 8.73$ (future collapse FC, magenta dash-dotted line), 
and $\Delta\rho = 5.41$
(zero gravity ZG, light blue solid line).
Tomato, blue, and light  blue vertical areas mark borders
of the HDC of the supercluster with $\Delta\rho \approx 40$, 
turnaround region of the supercluster 
main body, and zero gravity region  \citep{2020A&A...641A.172E}.  
\label{a2142zones}}
\end{figure}

From the observational side, the supercluster A~2142 is an example
of a collapsing supercluster with one high-density core 
\citep{2015A&A...580A..69E, 2020A&A...641A.172E}. 
As an example,  
we show in Figure~\ref{a2142sky}  the sky distribution of galaxies, groups and filaments
 in this supercluster
and in the cocoon around it, where we mark  the turnaround, future collapse (FC),
and zero gravity (ZG) regions of the supercluster. 
Figure~\ref{a2142zones} presents the density contrast versus
supercluster-centric distance for this supercluster, calculated using the distribution
of mass in SCl~A2142~\citep{2020A&A...641A.172E}. The  sizes of turnaround, future 
collapse, and zero gravity  regions are calculated based on the density contrast (mass distribution)
in the supercluster and the spherical
collapse model. For the Corona Borealis superclusters
these regions have been found by \citep{2021A&A...649A..51E}.

\begin{figure}[H]\vspace{-6pt}
%\isPreprints{\centering}{} % Only used for preprints
\includegraphics[width=10 cm]{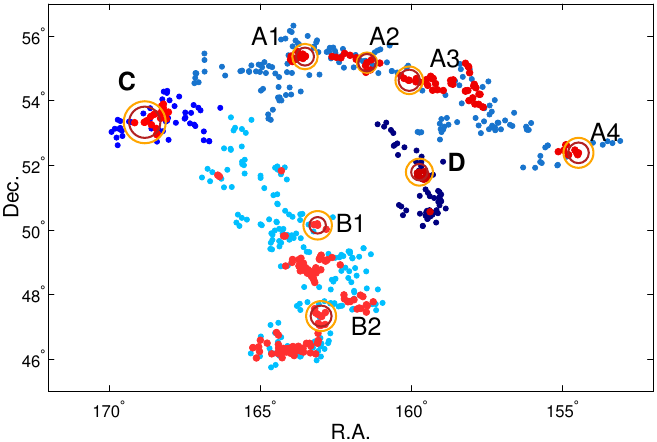}
\caption{Distribution of galaxies in the BGW superclusters in the sky plane. 
Red dots denote galaxies in the  HDCs of each supercluster, and 
blue dots show galaxies in the outskirts. Different shades of red and blue 
correspond to different BGW superclusters.
Dark red circles show the turnaround regions in each HDC, and orange circles show the future collapse 
regions.
Numbers with labels denote HDCs in each BGW supercluster. Figure by ME   \citep{2021A&A...649A..51E}.  
\label{bgwsky}}
\end{figure}

If a supercluster has several HDCs which are far apart from each other,
then each core may collapse separately in the distant future, and the
supercluster may split and form several smaller superclusters.
Such scenario has been predicted for the superclusters in the Sloan Great Wall
and in the BOSS Great Wall (BGW) \citep{2016A&A...595A..70E, 2021A&A...649A..51E}. 
We show high-density cores of
superclusters in the BGW with their turnaround and future collapse regions in Figure~\ref{bgwsky}. 
The sizes of these regions are calculated using the mass distribution
around the most massive cluster in each core.
In the BGW the distances between cores are so large that they will collapse separately
in the future and form smaller superclusters.

\begin{figure}[H]
%\isPreprints{\centering}{} % Only used for preprints
\includegraphics[width=10 cm]{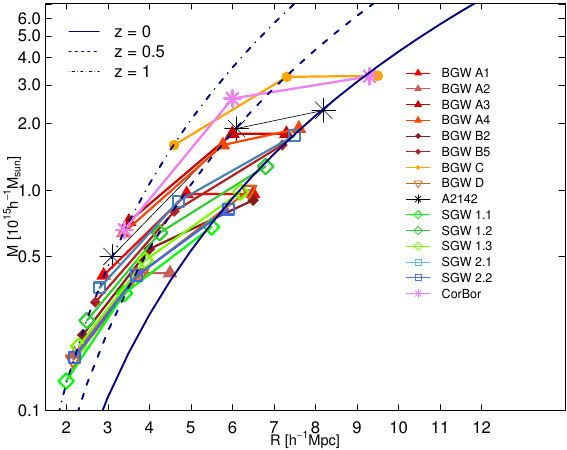}
\caption{Future collapse mass $M$ versus the radius of the HDCs $R$ in 
the Corona Borealis supercluster, in the A~2142 supercluster,
in the SGW superclusters and in the BGW superclusters
at redshifts $z = 0$ (right points), $z = 0.5$ (middle points), and
$z = 1$ (left points), as listed in the legend.
Black lines show theoretical mass--radius relation from the spherical collapse model for redshifts
$z = 0.0$ (solid line), $z = 0.5$ (dashed line), and $z = 1.0$ (dot-dashed line)
 \citep{2021A&A...649A..51E}.  
\label{cores}}
\end{figure}

If the distance between rich clusters in the HDCs of superclusters 
is smaller than the future collapse radius, then during the future evolution
of a supercluster, such HDCs may merge and form a massive HDC. 
This evolution scenario was predicted, for example,
for the Corona Borealis supercluster and for the Shapley supercluster
in the nearby Universe~\citep{2021A&A...649A..51E, 2015A&A...575L..14C}. 
A detailed comparison of the radii and masses of superclusters in the nearby
Universe and at redshift $z = 0.5$ was presented in~\citep{2021A&A...649A..51E},
and we show this comparison also in Figure~\ref{cores}.
In this figure, the masses and sizes of nearby superclusters (redshift 
$z = 0$) are taken from~\citep{2016A&A...595A..70E, 2020A&A...641A.172E, 2021A&A...649A..51E}, and
extrapolated to higher redshifts based on the predictions of the spherical
collapse model. In the case of the BOSS Great Wall (BGW), superclusters at redshifts $z = 0.46$, and
the masses and sizes
of collapsing regions are based on these values at the BGW redshift, 
$z \approx 0.5$, and extrapolated to lower and higher redshifts using
the spherical collapse model and the distribution of mass in the BGW
superclusters.
This figure shows that the most massive high-density core among the BGW superclusters,
BGW C, has a mass and size similar to the core of the Corona Borealis supercluster
in the nearby Universe, of the order of $M \approx$ 3--$4 \times~10^{15}M_\odot$ 
and 10~\Mpc\ at redshift $z = 0$.
The masses and sizes of the collapsing cores of other superclusters
lie in the range of\mbox{ $M \approx$ 0.2--$2 \times 10^{15}M_\odot$}
and 2--8~\Mpc.
We also refer to \citep{2024PASA...41...78Z}, which provides an analysis of HDCs
(nucleation regions of superclusters) and their future evolution.
Ref. \citep{2021A&A...649A..51E} proposed that the number and properties of collapsing cores of superclusters
may serve as a \mbox{cosmological test.}

The reliability of the predictions made for observed superclusters,
depends, of course, on how well the masses
of turnaround regions of HDCs are estimated. For example, ref.~\citep{2021A&A...649A..51E} estimated
that in order to merge in the distant future, the masses of different HDCs
in the BGW should be ten times higher than estimated in their study,
which leads to the mass-to-light ratios $M/L \approx 3000$. This value 
is unrealistic. Therefore, ref. \citep{2021A&A...649A..51E} concluded that the cores of the
BGW superclusters will separate in the distant future.

\begin{figure}[H]
%\isPreprints{\centering}{} % Only used for preprints
\includegraphics[width=12 cm]{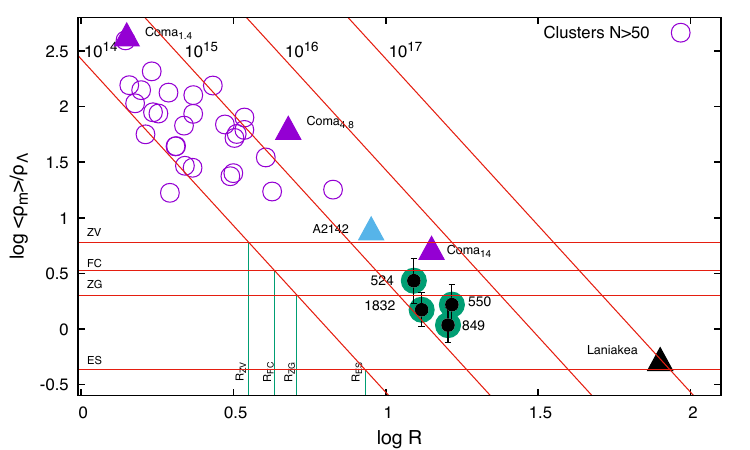}
\caption{ $\Lambda$ {significance} %MDPI: Please change the hyphen (-) into a minus sign (−, “U+2212”) in the figure, e.g., “-1” should be “−1”.
 diagram for SDSS clusters of galaxies (purple) 
with quasispherical superclusters marked in green. The Laniakea supercluster, 
the central regions of the supercluster A2142 and the Coma cluster 
with three different mass and size estimates are shown. 
Inclined lines show different mass limits. 
Green vertical lines show the values of the $R_{\rm ZV}$  (zero velocity radius), 
$R_{\rm ZG}$ (zero gravity radius), and the $R_{\rm ES}$ 
(the Einstein--Straus radius for the mass {$10^{14} \,M_{\odot}$}) 
and the horizontal lines show logarithmic values for the constant 
ratios of the 
$\langle \rho_{\mathrm{M}}\rangle / \rho_{\Lambda}$ = 6 ($\rm ZV$), 2 ($\rm ZG$), and 3/7 ($\rm ES$). 
Figure by Pekka Heinämäki.  
\label{lsig}}
\end{figure}

Finally, we present Figure~\ref{lsig}, which shows the dynamical state of galaxy clusters and
superclusters from the dark energy (DE) (and dark matter) perspective, the so-called
$\Lambda$ significance diagram  \citep{2015A&A...577A.144T, 2022A&A...668A..37H}.
The energy density ratio
$\langle \rho_{\mathrm{M}} \rangle / \rho_{\Lambda}$, as calculated for
the system under study,  measures the influence of DE. The $\Lambda$ significance diagram is a 
 $log(\langle \rho_{\mathrm{M}} \rangle / \rho_{\Lambda}) $ versus $\log R$ graph, where  $R$ is the
radius of a system, $\langle \rho_{\mathrm{M}}\rangle$ is its average mass density, and
$\rho_{\Lambda}$ is the DE density equal to the global value 
$\rho_{\Lambda} \approx 6 \times 10^{-30}$ g cm$^{-3}$.

The location of a galaxy system in the 
$\Lambda$ significance diagram
indicates whether its overall dynamics is dominated by gravity or by the outward
`antigravity' expulsion of DE.
The lines in the figure are as follows.
The ratio $\langle \rho_{\mathrm{M}}\rangle / \rho_{\Lambda}$ can be expressed as
$\log (\langle \rho_{\mathrm{M}}\rangle / \rho_{\Lambda}) = 0.43 + \log (M/10^{12}M_\odot)
- 3 \times \log (R/\mathrm{Mpc})$,
where $M$ is the mass within the radius $R$ of a system.
The intersections of the lines with horizontal lines of constant
$log(\langle \rho_{\mathrm{M}}\rangle / \rho_{\Lambda})$ give the following scales or radii for the 
mass $M$:
\begin{enumerate}
\item The zero-velocity radius $R_{\rm ZV}$ (at this distance, a test particle system-centric radial velocity $u=Hr-v_{pec}$ is $v_{pec}=Hr$ and $u=0$). This corresponds to the turnaround radius in the spherical collapse model.
\item The zero-gravity radius $R_{\rm ZG}$ (gravity force equal to Einstein’s  antigravity force). Thus, acceleration is zero, $du/dt=0$,  which indicates the
 maximum 
radius of a gravitationally bound system at the present epoch. 
\item The Einstein--Straus radius $R_{\rm ES}$ (the radial velocity reaches the Hubble velocity, \mbox{$u=Hr$)}. This last, longest distance corresponds to the spherical volume where the mass $M$ produces an average density that is equal to the cosmic global density. 
\end{enumerate}

In Figure ~\ref{lsig}, these three different radii are shown for the systems of mass $10^{14} \,M_{\odot}$.
For spherical and homogenous systems, the horizontal lines ${\rm ZV}$, 
 ${\rm ZG,}$ and ${\rm ES}$ correspond to the 
constant ratios of $\langle \rho_{\mathrm{M}}\rangle / \rho_{\Lambda}$ = 6, 2, and 3/7. 

In Figure~\ref{lsig}, clusters of galaxies (from the SDSS survey) are shown with purple circles; 
they all lie in the region
of structures that have pass turnaround and are collapsing, as one could expect for clusters.
%EE: Please check that the intended meaning has been retained  %ME: OK
Based on the density distribution, the locations of several clusters and superclusters
in this figure is shown. The supercluster A2142 core 
 region is shown as a blue triangle near the ZG line. For the Coma cluster, regions of several radii are presented. 
At small radius ($Coma_{1.4}$),  the figure shows the location of the Coma cluster itself,
larger radius ($Coma_{4.8}$) corresponds to the Coma cluster with its region
of influence, in a good agreement with recent estimates using Cosmicflows-4 data 
\citep{2025arXiv250404135B}. The largest radius ($Coma_{14}$) shows the FC region around the Coma cluster.
 The Laniakea supercluster BoA borders
 are located near the $\langle \rho_{\mathrm{M}}\rangle / \rho_{\Lambda}$ = 3/7 ($\rm ES$) line.
Only the central part of Laniakea may collapse in the future \citep{2015A&A...575L..14C}.  
Almost spherical superclusters (objects 524, 550, 849, and 1832 in the figure \citep{2022A&A...668A..37H}) lie between the core of SCl~A2142
and the Laniakea, showing that their outskirts are not collapsing yet.

\section{Superclusters as an Environment for Galaxies, Groups, and Clusters}
\label{sect:gal} 

Already early studies of superclusters provided evidence of the
coevolution of superclusters, galaxies, and galaxy systems embedded in them
\citep{1980MNRAS.193..353E}.
Galaxies can be divided into two major classes according to their 
star formation and morphological properties: quiescent galaxies, which are typically red and
of early-type, and actively star-forming galaxies having blue colours and late types
\citep{2001AJ....122.1861S, Bell:2017vy, Bluck:2020vt, 2022MNRAS.513..439D}.
The analysis  of galaxy populations in the Perseus-Pisces supercluster
demonstrated that the central parts of this supercluster are mostly populated
by early-type galaxies, while the percentage of late-type galaxies increases in the 
supercluster outskirts
\citep{1978MNRAS.185..357J, 1980MNRAS.193..353E, 1986ApJ...300...77G}. 
This is now  known as the morphological segregation or the morphology--density
relation, in which 
`morphology' denotes various properties of galaxies:
their morphological type, star formation activity, colour, and so on. This relation tells
that rich galaxy clusters and especially their central parts are
mostly populated by early-type galaxies, and late-type galaxies are 
preferentially located on the outskirts of clusters or in poor groups 
\citep{1980ApJ...236..351D}. In fact,
the morphology--density relation extends from central parts of galaxy clusters up to the lowest global-density
environment, populated by very poor galaxy groups and single galaxies 
\citep{1987MNRAS.226..543E, 2022A&A...668A..69E}. Early-type galaxies are located
in the central part of rich clusters or in central parts of poor groups in filaments
between clusters   \citep{1988MNRAS.234...37E}.

\begin{figure}[H]
%\isPreprints{\centering}{} % Only used for preprints
\includegraphics[width=10 cm]{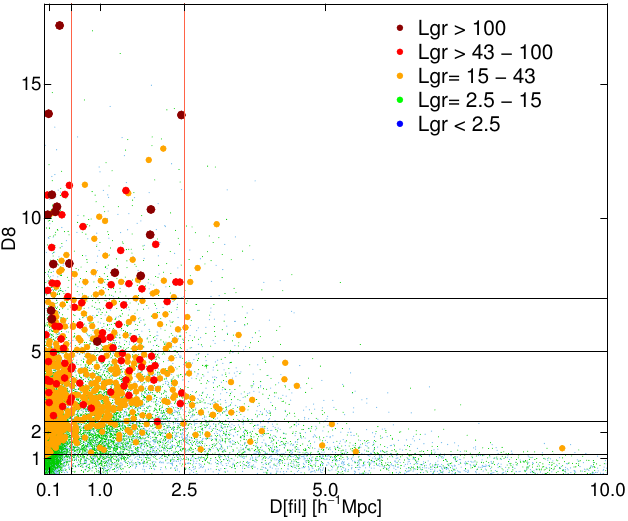}
\caption{Global luminosity-density $D8$ versus distance from the nearest filament axis 
$D_{fil}$ for galaxy groups and clusters of various luminosity.
Dark red circles refer to groups with luminosity $L_{gr} \geq 100 \times10^{10} h^{-2} L_\odot$,
red circles- groups with luminosity $43 \leq L_{gr} \leq 100 \times10^{10} h^{-2} L_\odot$,
orange circles- groups with luminosity $15 \leq L_{gr} \leq 43 \times10^{10} h^{-2} L_\odot$,
green circles- groups with luminosity $2.5 \leq L_{gr} \leq 15 \times10^{10} h^{-2} L_\odot$,
and blue circles- groups with luminosity $L_{gr} \leq 2.5 \times10^{10} h^{-2} L_\odot$.
Note that all the most luminous clusters lie in superclusters with $D8 \geq 5$.
Figure by ME \citep{2024A&A...681A..91E}.  
\label{d8dfil}}
\end{figure}

Recent studies have shown that the global environment strongly affects the richness of galaxy groups and clusters.
X-ray clusters are more frequent in superclusters~\citep{2001AJ....122.2222E}.
Also, clusters in superclusters have a higher probability
to have substructure than isolated clusters~\citep{2012A&A...542A..36E}.
In the Ursa Major supercluster, relaxed
groups lie in high-density regions, while non-relaxed clusters with substructure
can preferentially be found in the outskirts of the supercluster \citep{2013A&A...551A.143K}.
Figure~\ref{d8dfil} shows the effect of the environment on the richness (luminosity)
of groups in the global luminosity-density $D8$--distance to the nearest
filament axis $D_{fil}$ plane \citep{2024A&A...681A..91E}. 
The 
luminosity--density values in this figure ($D8$, calculated using  smoothing length $8$~\Mpc)
are expressed in units of mean luminosity density, \mbox{$\ell_{\mathrm{mean}}$ =  
1.65 $\times~10^{-2}$ $\frac{10^{10} h^{-2} L_\odot}{(\vmh)^3}$ \citep{2012A&A...539A..80L, 2022A&A...668A..69E}. 
In Figure~\ref{d8dfil}}, galaxy groups and clusters can be considered as being located in filaments or in filament outskirts
if the distance to the nearest filament axis $D_{fil} \leq 2.5$~\Mpc\ ($D_{fil} \leq 0.5$~\Mpc\
for filament members).
In the global luminosity-density field, the density threshold $D8 = 5$ marks superclusters,
and $D8 = 7$ corresponds to the HDCs of superclusters. 

There are several important aspects to notice in this figure.
It shows that the richest and most luminous clusters with luminosity
$L_{gr} \geq 100 \times10^{10} h^{-2} L_\odot$ all lie in superclusters or in the HDCs of
superclusters  with $D8 \geq 5$ in filaments or filament outskirts, with $D_{fil} \leq 2.5$~\Mpc. 
Clusters grow by infall of groups and
galaxies along filaments, and a dense environment of superclusters is especially suited for that.
Refs. \citep{2003A&A...401..851E, 2005A&A...436...17E} called this an environmental enhancement of galaxy groups near rich
clusters. 
One can also notice that intermediate luminosity groups with 
$43 \leq L_{gr} \leq 100 \times10^{10}~h^{-2} L_\odot$ and 
$15 \leq L_{gr} \leq 43 \times10^{10}~h^{-2} L_\odot$
are in abundance at the outskirts of superclusters ($1 < D8 < 5$). Such groups avoid the lowest global
density regions (watersheds) between superclusters, with $D8 < 1$. 
However, very poor groups (mainly galaxy pairs and triplets with luminosity
$L_{gr} \leq 2.5 \times10^{10}~h^{-2} L_\odot$) and galaxies which do not belong to any group 
(single galaxies)
can be found everywhere in the cosmic web, from superclusters to the lowest global
density environments (voids or watershed regions) \citep{2022A&A...668A..69E, 2024A&A...681A..91E}.
We note that single galaxies and very poor groups with $D_{fil} > 2.5$~\Mpc\ are, 
most probably, members of faint filaments which were not detected based on data from the
SDSS MAIN spectroscopic survey, used in \citep{2024A&A...681A..91E}.
The change in group and cluster properties at threshold densities $D8 = 5$
which have been used to define superclusters, means that when looking for the physical definition
of superclusters, the analysis of the dependence of the properties of galaxy groups and clusters
on the global density may provide additional information.
Rich clusters are located in superclusters, and this
is related to the high value of their correlation amplitude \citep{1996astro.ph.11148B}.

A possible connection between the location of galaxy groups and clusters
and their properties is known as a halo assembly bias, which addresses the question
whether other factors are important in shaping the properties of groups and 
clusters apart from their masses~\citep{2018ARA&A..56..435W}. This problem has been addressed in many studies.
These studies have shown that clusters in high-density regions (cosmic web nodes) tend 
to more often be X-ray sources, and to have 
higher X-ray luminosities and temperatures 
than groups and clusters in lower-density environments, such as filaments
\citep{2021MNRAS.500.1953M, 2024MNRAS.527..895P, 2024A&A...688A.186L}.
This may mean that clusters in denser environments, such as superclusters, have
experienced more mergers than isolated clusters in their assembling history
\citep{2023A&A...676A.127D, 2024A&A...688A.186L}.

\begin{figure}[H]
%\isPreprints{\centering}{} % Only used for preprints
\includegraphics[width=8 cm]{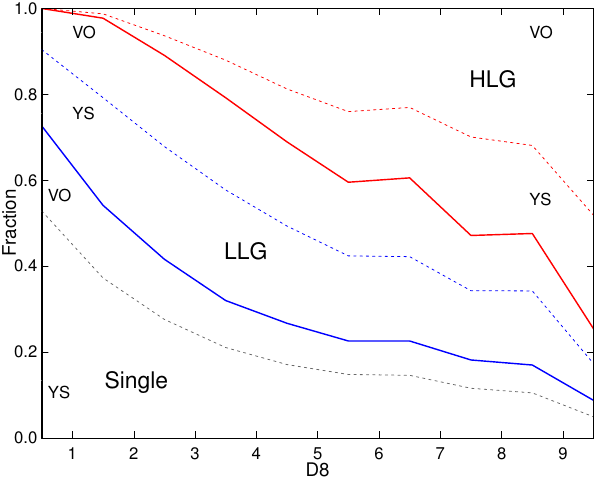}
\caption{Fractions of high- and low-luminosity groups and single galaxies 
(HLG, LLG, and Single, correspondingly) 
(divided according to the star formation properties of galaxies in
groups and among single galaxies) in regions of various global luminosity-density $D8$.
Global densities \mbox{$D8 \geq 5$} (in units of mean luminosity-density) correspond to superclusters,
and $D8 \leq 1$ to the lowest density regions between superclusters (watersheds).
HLG refers to high-luminosity groups with \mbox{$L_{gr} \geq 15 \times10^{10} h^{-2} L_\odot$,}
LLG to groups with lower luminosity, $L_{gr} < 15 \times10^{10} h^{-2} L_\odot$, and single to galaxies
which do not belong to any group. VO and YS indicate galaxy populations in groups.
VO refers to galaxies with very old stellar populations with  $D_n(4000)$ index
 $D_n(4000) \geq 1.75$, and YS to galaxies with young stellar
populations with  $D_n(4000) < 1.75$.
In each global density interval ($D8 = 0--1, 1--2$ and so on),
the sum of fractions is $F_{single, VO} + F_{single, YS} + F_{LLG, VO} + F_{LLG, YS} + 
F_{HLG, VO} + F_{HLG, YS} = 1$. Figure by ME \citep{2022A&A...668A..69E}.
\label{frac}}
\end{figure}

 The star formation properties of galaxies are less affected by the global environment.  
We show this in Figure~\ref{frac},
which presents the group and galaxy content of various global luminosity-density 
regions based on the Sloan Digital Sky survey data. 
As in Figure~\ref{d8dfil}, one can see that in the supercluster environment, 
the abundance of luminous groups and clusters increases. Poor 
groups and single galaxies lie preferentially in the low global density environment.
The galaxy and group populations of superclusters differ from those
in a low-density environment between superclusters. The percentage of red, quiescent galaxies
in superclusters is higher than in low-density environments, up to 60\% \citep{2007A&A...464..815E, 2012A&A...545A.104L}.
In the lowest global density environment
populated by single galaxies and very poor groups, up to one-third of all galaxies
are quiescent \citep{2022A&A...668A..69E}. 

Groups of the same richness in superclusters are more luminous than those in the low-density
regions. This phenomenon is especially strong for poor groups.
Moreover, groups in superclusters embed higher percentages of red,
early-type galaxies  in comparison with groups of the same richness in low-density
regions  \citep{2012A&A...545A.104L, 2014A&A...562A..87E, 2022A&A...668A..69E,
2024A&A...681A..91E}. 
Even single galaxies, from the volume-limited samples of the same absolute magnitude limit
in superclusters, have, on average, higher luminosities than 
single galaxies elsewhere \citep{2024A&A...681A..91E}.
%EE: Please check that the intended meaning has been retained  %ME  OK
Also, in superclusters of filament morphology, 
galaxy groups of the same richness host galaxies with larger stellar masses, and 
a higher fraction of early-type and red galaxies in comparison with groups in superclusters
of spider morphology \citep{2014A&A...562A..87E}.

Group richness varies strongly with the global environment. 
In superclusters quiescent galaxies lie preferentially in rich groups and clusters.
In other regions of the cosmic web, quiescent galaxies are hosted by poor groups, or they do
not belong to any group \citep{2022A&A...668A..69E}. Considering that superclusters host
approximately 15\% of all galaxies, one can see that quiescent galaxies in low-density regions
outnumber, by far, such galaxies in rich groups in superclusters.
Even among dwarf galaxies in low-density  
environments, there are quiescent galaxies \citep{1988MNRAS.234...37E, 2025A&A...693L..16B}.
These results demonstrate that in quenching the star formation in galaxies, 
processes, which are effective within and in the neighbourhood of galaxies and their dark matter 
haloes, play an important role, and the effect of larger-scale environments is less
important \citep{2022A&A...668A..69E, 2025A&A...693L..16B}.
%EE: Please check that the intended meaning has been retained %ME: made minor changes

The largest difference in galaxy properties within superclusters and elsewhere
in the cosmic web is related to the properties of the brightest galaxies in groups and 
clusters. The brightest galaxies of the richest clusters (BCGs), which lie in superclusters, with luminosities
$L_{gr} \geq 15 \times10^{10} h^{-2} L_\odot$, are mostly quiescent,
with stellar populations older than at least $4$~Gyrs. 
The BCGs have stellar masses $log M^* > 11.0$, red colours, and stellar velocity dispersions
$\sigma^* > 300$~$km/s$ \citep{2015MNRAS.448.1483L, 2020ApJ...891..129S, 2022ApJ...931...31S,
2021MNRAS.507.5780M, 2022A&A...668A..69E, 2024A&A...681A..91E}.
These properties suggest that the formation and growth of the present-day BCGs began at early stages
of their host cluster's evolution.
The poor groups,
which can be found everywhere in the cosmic web, may also have star-forming
brightest galaxies (\citep{2022A&A...668A..69E, 2022A&A...665A..44A, 2024A&A...681A..91E}
and references therein).

We may assume that individual rich superclusters, with thousands of galaxies (in volume-limited
samples, which include brighter galaxies only) have, on average, similar galaxy populations.
Still,  surprisingly, the galaxy populations of individual rich superclusters
are different \citep{2011ApJ...736...51E}. The comparison of galaxy populations in elongated and 
flattened superclusters shows that
superclusters of different overall shape have similar stellar populations \citep{2013MNRAS.428..906C}.
Therefore, these differences are related to the inner structure
of superclusters. Filament-type superclusters, with a small number of filaments between clusters,
embed, on average, a higher percentage of red, quiescent galaxies than spider-type 
superclusters, in which clusters are connected by a large number
of filaments \citep{2014A&A...562A..87E, 2017ApJ...835...56C}.
In spider-type superclusters, galaxy clusters have
larger amount of substructure, and this affects the star-formation properties of 
galaxies in {them}%MDPI: There are two refs 168 citation, we removed the repeated, please confirm.
~\citep{2017ApJ...835...56C}.
At redshifts around $z \leq 0.9$, more
connected clusters tend to have a higher number of 
star-forming member galaxies (web-feeding model) 
\citep{2019MNRAS.490..135L, 2024ApJ...976..154K}. This result is in agreement
with the earlier result that superclusters with dynamically active clusters
have a higher percentage of star-forming galaxies.

The properties of {\it filaments} in superclusters have been analysed 
in several studies. 
The studies of the Local supercluster have revealed the filamentary pattern
around the Virgo cluster \citep{1982ApJ...257..389T, 2022A&A...657A...9C}.
Early analysis of the filamentary patterns in and around superclusters was performed
mainly using percolation (clustering) analysis.
At present, several filament finders are available, as the 
Discrete Persistent Structure Extractor (DisPerSE), Nexus
(an algorithm for the identification of cosmic web environments: clusters, 
filaments, walls, and voids), a graph-based filament detection method
T-Rex, a marked point process (Bisous model) based filaments, and others 
\citep{2011MNRAS.414..350S, 2013MNRAS.429.1286C, 2014MNRAS.438.3465T, 2016A&C....16...17T,
2020A&A...634A..30M, 2020A&A...637A..31S, 2020A&A...637A..18B}.

The number of filaments connected to a cluster defines the cluster's
connectivity $C$. Three filaments connected to the Coma cluster
were found with the DisPerSE filament finder by \citep{2020A&A...634A..30M}.
Applying Bisous filaments, the connectivity of the richest clusters in the
A~2142 and in the Corona Borealis superclusters have been found to be\mbox{ $C$ = 2--6
\citep{2020A&A...641A.172E, 2021A&A...649A..51E}.}
First detections of the connectivity of superclusters have shown 
that the connectivity of the A~2142, the Corona Borealis and the Shapley
superclusters is in the range of \mbox{$C$ = 6--7 \citep{2020A&A...641A.172E, 2021A&A...649A..51E,
2024A&A...689A.332A}}. The connectivity of poor groups outside superclusters
is lower, usually in the range of $C$ = 1--2. 

Typically, filaments within superclusters are short, with a length less than $5$~\Mpc, but
long filaments in superclusters may have lengths larger than $15$~\Mpc, and such
filaments may extend farther away from superclusters, to low-density regions
around \mbox{them \citep{2020A&A...641A.172E, 2020A&A...637A..31S, 2021A&A...649A..51E}.}
Galaxies closer to the filament axis are more massive than those farther away from filaments
\citep{2020A&A...637A..31S}.

Usually, filaments have been traced by galaxies. In the study of the
Shapley supercluster filaments, determined with the T-REX filament finder,
 were analysed also using data on gas \citep{2024A&A...689A.332A}.
Gas content in filaments between clusters in superclusters is one factor that affects
the star-formation properties of galaxies. This has been demonstrated for the richest
supercluster in the nearby Universe, the Shapley supercluster \citep{2024A&A...689A.332A}.

High-redshift protoclusters are candidates of the present-day rich clusters that lie in
superclusters. Thus, the study of protoclusters and their galaxy content provides information
on the galaxy evolution in superclusters. For example, quenched galaxies have been 
detected in a protocluster at redshift $z = 3.37$ \citep{2022ApJ...926...37M}.
Quenched galaxies with stellar populations older than $10$~Gyrs (galaxies
which have stopped forming stars at least $10$~Gyrs ago) have been found among both
the brightest cluster galaxies and satellites in rich clusters in superclusters
in the nearby Universe \citep{2022A&A...668A..69E}.

%%%%%%%%%%%%%%%%%%%%%%%%%%%%%%%%%%%%%%%%%%%%%%%

The important and fast-developing field, which provides information on the coevolution of galaxies and structures
in which these are embedded, is the study of {\it alignments} between galaxies, clusters,
and their host systems, including superclusters \citep{1980MNRAS.193..353E, 1989ApJ...347..610W,
2002ASSL..276..299P, 2004ogci.conf...19P, 2015SSRv..193....1J} (and references therein).
Refs. \citep{1978MNRAS.185..357J, 1980MNRAS.193..353E} showed that in the Perseus-Pisces 
supercluster, the brightest galaxies in groups and clusters, as well as groups and clusters
themselves, are aligned along the supercluster axis, which is a signature of their
coevolution in the cosmic web. Correlated alignments between neighbouring groups and clusters
in superclusters and in the cosmic web up to scales $100$~\Mpc\ were detected by \citep{1989ApJ...347..610W,
2002ASSL..276..299P, 2002MNRAS.329L..47P, 2004ogci.conf...19P, Paz2011}.
In the supercluster SCl~A2142,
the brightest galaxies in the main cluster of the supercluster, A2142, are 
aligned with the cluster A2142. Also, the radio and X-ray haloes of this cluster
are elongated and aligned along the cluster and supercluster axis, as 
are groups infalling into the cluster. This suggests that structures in the
supercluster 
have grown through merging and accretion along the supercluster axis.
Moreover,  SCl~A2142 hosts an FRII radio
galaxy with giant lobes, which are also aligned with the supercluster axis 
\citep{2018A&A...620A.149E}.

One interesting problem is the effect of the supercluster environment on the properties
of radio galaxies, especially to radio galaxies with giant $\geq 3$~Mpc radio lobes. Studies of 
giant radio galaxies (GRGs)  in various environments show that radio galaxies with radio lobes 
$\geq 3$~Mpc preferentially lie in low-density environments \citep{2024A&A...687L...8S}.
Moreover, identifying GRGs
with polarised emissions behind superclusters can be used to estimate the magnetic field strength of
sub-micro Gauss  within the supercluster environment \citep{2024A&A...687L...8S}.

\section{Superclusters and the Cosmic Microwave Background (CMB)}
\label{sect:cmb}

\begin{figure}[H]
%\isPreprints{\centering}{} % Only used for preprints
\includegraphics[width=12 cm]{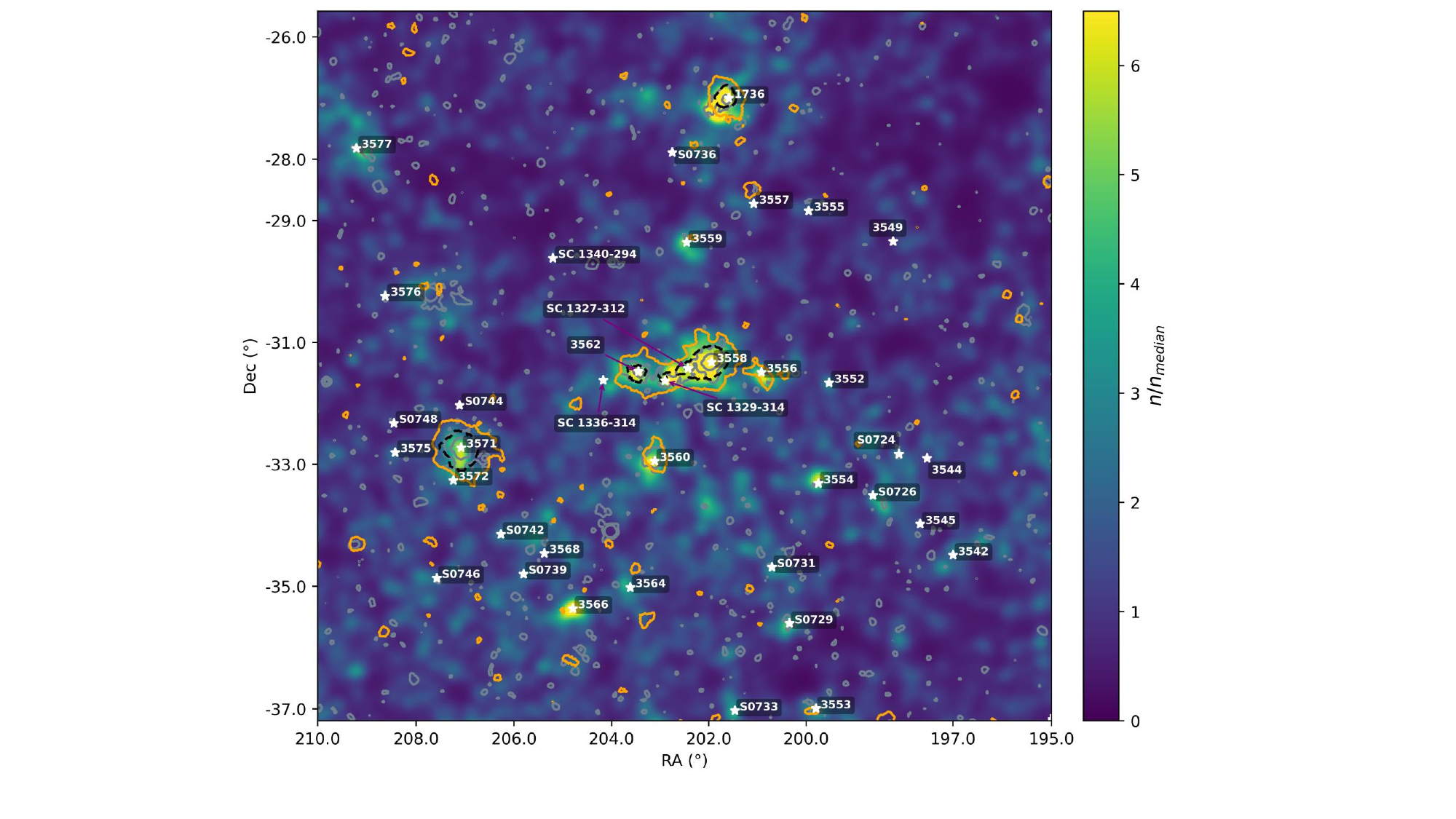}
\caption{{Central} %MDPI: Please change the hyphen (-) into a minus sign (−, “U+2212”) in the figure, e.g., “-1” should be “−1”.
 part of the Shapley Supercluster
within a redshift range of $0.03<z<0.083$. White stars 
mark galaxy clusters from \citep{2006A&A...447..133P}. 
The colour scale represents the galaxy number density, 
normalised by the median density in each pixel. The outer orange solid contours and inner 
black dashed contours correspond to Compton $y$ parameter values of 
$4.4\times 10^{-6}$ and $1.1\times 10^{-5}$, respectively, in the tSZ 
reconstructed map from the \textit{Planck} all-sky survey. 
The outer and inner grey contours correspond to the X-ray signal of 2000 
counts$\,$s$^{-1}\,$arcmin$^{-2}$ and $10^{4}$ counts$\,$s$^{-1}\,$arcmin$^{-2}$, 
respectively, in the ROSAT all-sky survey map.
 Figure from \citep{2024A&A...689A.332A}.
\label{shapsz}}
\end{figure}

High-precision measurements of the CMB with the WMAP and especially {\it Planck} satellites
opened a new window into the many branches of cosmology, including supercluster studies. 
First of all, I mention the discovery of a new supercluster at redshift $z = 0.45$ in the distribution of 
galaxy clusters detected by analysing the XMM-Newton follow-up observations
for the confirmation of Planck cluster catalogue \citep{2011A&A...536A...9P}.
The hot intercluster gas in superclusters can be detected from  
the thermal Sunyaev--Zeldovich (tSZ) signal using the full-sky Compton parameter
maps (y-maps) reconstructed 
from the Planck multi-frequency channel maps \citep{1970Ap&SS...7....3S, 2022MNRAS.509..300T}.
This approach has been applied, for example, by \citep{2022A&A...668A..37H, 2024A&A...689A.332A}.
In particular, in this way, ref. \citep{2024A&A...689A.332A} investigated
the relation between the star-formation rate in galaxies and the
Compton y parameter, and its dependence on the environment defined by intercluster filaments in the Shapley
supercluster. 
 Figure~\ref{shapsz} shows the galaxy density maps of the central part of the Shapley
supercluster, in which optical clusters from~\citep{2006A&A...447..133P} are plotted with
stars, and contours of different colour correspond to Compton $y$ parameter values
and X-ray contours from the ROSAT all-sky survey map, as described in the \mbox{figure caption.}

The largest structures in the cosmic web---superclusters and supervoids
affect CMB photons, very slightly heating or cooling them, correspondingly.
This effect is known as the
integrated Sachs--Wolfe (ISW) effect (see \citep{2008ApJ...683L..99G} for details and references).
This effect has been detected
in the studies of superclusters and supervoids  using various datasets,
including Dark Energy Survey data, and represents one interesting direction for the future
studies 
 \citep{2008ApJ...683L..99G, 2016ApJ...830L..19N, 2022MNRAS.515.4417K, 2025arXiv250401669D}.
Ref. \citep{2025arXiv250119236B} described several superstructures of X-ray clusters in
the nearby Universe and mention that these superstructures should produce a modification on the 
CMB background.

\section{The Rich and Important Superclusters }
\label{sect:rich} 

For us in the Milky Way galaxy, the most important supercluster is, of course, our cosmic
home, {\it  the Virgo supercluster}. Enhanced density of nebulae in the direction of the Virgo 
constellation was noticed already by William and John Herscel \citep{2006Ap&SS.302...43H, Juhan_thesis}.
Gerard de Vaucouleur noticed a concentration of galaxy systems in Virgo,
and described this as a ``supergalaxy'' or
``supercluster'' \citep{1953AJ.....58...30D, 1956VA......2.1584D}. 
The Virgo supercluster consists of the main cluster---the Virgo cluster, and a filamentary
patterns around it, with the extent approximately
20~\Mpc\ \citep{1982ApJ...257..389T, 1984MNRAS.206..529E, 1988MNRAS.234...37E, e07, 2022A&A...657A...9C}.
Our Local Group of galaxies is located in the galaxy filament which connects the Virgo and
the Fornax superclusters  \citep{2024A&A...690A..92R}.
The studies of the Local supercluster are important, as in this region it is possible to study
much 
fainter galaxies than in studies of more distant superclusters. One interesting result
of such studies is the discovery of relatively isolated, early-type dwarf
galaxies \citep{1988MNRAS.234...37E, 2025A&A...693L..16B}. These studies extend the quenching of star formation
in galaxies to much lower luminosities than in more distant samples.

\begin{figure}[H]
%\isPreprints{\centering}{} % Only used for preprints
\includegraphics[width=12 cm]{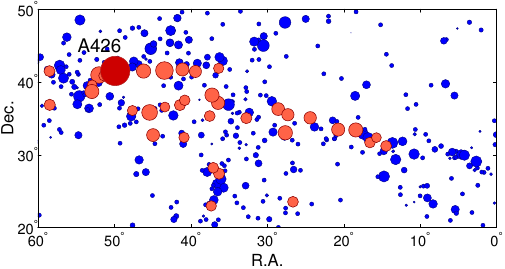}
\caption{Distribution of galaxy groups from \citep{2016A&A...588A..14T}
 in the Perseus-Pisces supercluster in the sky plane. 
Red circles: high-mass clusters with mass $M \geq 10^{14}M_\odot$
in the supercluster region. Dark red circle: Abell cluster A~426.
Blue circles: lower mass groups and clusters. Figure by ME.
\label{pers}}
\end{figure}

The very elongated {\it  Perseus-Pisces supercluster} or the Perseus-Pisces chain,
plotted in Figure~\ref{pers},  is one of those
structures which led to the discovery of the cosmic \mbox{web \citep{1978MNRAS.185..357J,
1980MNRAS.193..353E}}. The Perseus-Pisces supercluster is also the first supercluster
for which it was found that  galaxy clusters in it are elongated along the
supercluster axis. %This is  shown in Figure~\ref{pers}. 
The brightest cluster galaxies are aligned with the supercluster axis as well.
In addition, these first studies already demonstrated
 the presence of the large-scale morphological segregation of galaxies along the Perseus-Pisces filaments  
\citep{1978MNRAS.185..357J, 1980MNRAS.193..353E, 1986ApJ...300...77G}.
These properties are signatures of the common origin and evolution of galaxies,
galaxy clusters, and the whole supercluster \citep{1978MNRAS.185..357J,
1980MNRAS.193..353E}.

The nearby {\it  {Coma supercluster}} is another supercluster that led to the discovery of the cosmic web
\citep{1984MNRAS.206..559T, 2020MNRAS.497..466S}. The concentration of nebulae in the direction of the
Coma constellation was first noticed by William Herschel in  1802 \citep{Juhan_thesis}.
The Coma supercluster with rich Coma (Abell A1656) and A1367 clusters is one of the best-studied superclusters~\citep{2020MNRAS.497..466S}. However,
there is no consensus whether in the distant future its two richest galaxy clusters,
A1656 and A1367, will merge, or if they are not massive enough to merge, and instead they will move away
from each other as the Universe expands \citep{2024PASA...41...78Z}.
 The Coma supercluster lies in a filament that connects our Local supercluster, the Coma, and the 
Hercules supercluster.

{\it  {The Hercules supercluster}}, at redshift $z \approx 0.04$, embeds 12
Abell clusters in \citep{2001AJ....122.2222E} supercluster catalogue.
This supercluster is very clumpy, and these clumps are often
considered separate, collapsing in the future  superclusters. One of them, centred at the cluster A2052,
is one of the most spherical superclusters \citep{2022A&A...668A..37H}.
The central cluster of another component of the Hercules supercluster,
A2199, is rotating \citep{2017ApJ...842...88S}. 
Kinematics of the Hercules supercluster determined by Abell clusters A2147, A2151, and A2152
were studied by \citep{1998AJ....115....6B}, who found that the Hercules supercluster,
traced by these clusters, may be gravitationally bound, with a mass of approximately 
$M \approx 7.6 \pm 2.0 \times~10^{15}h^{-1}M_\odot$. Ref. 
\citep{2022MNRAS.509.3470M} analysed in detail the kinematics  
of the Hercules supercluster, traced by Abell clusters A2151, A2152, and A2147 in their study.
They estimated that the total mass of the supercluster is
$M \approx 2.1\times~10^{15}M_\odot$; this is the lower mass limit for the full supercluster.

The richest supercluster in the nearby Universe is {\it  {the Shapley supercluster}}
 (Figure~\ref{shapsz};  see \citep{2024A&A...689A.332A} for detailed description and references).
In the core region of the Shapley, multiple filaments of galaxies and gas connect galaxy
clusters. The role of filaments in quenching galaxies in them was recently discussed
in detail by \citep{2024A&A...689A.332A}. 
Moreover, $\approx 7$~\Mpc\ long filament of hot gas between cluster pairs A3530/32 and A3528N/S  in this supercluster
was detected by \citep{2025A&A...698A.270M}, combining optical and X-ray observations.
The mass of the Shapley supercluster
is $M \approx 5\times 10^{16}~h^{-1}M_\odot$ \citep{2006A&A...447..133P}.
The core region of the Shapley supercluster with several merging rich clusters,
X-ray clusters among them, is collapsing,  and according to the mass and size
estimations, it will form the
most massive collapsing structure in the nearby Universe detected so far,
with a mass and radius  of the order of  $M \approx 1.3\times 10^{16}~h^{-1}M_\odot$
and $R \approx 12.4$~\Mpc\  \citep{2000AJ....120..523R, 2015A&A...575L..14C}.

The Coma, Hercules, and Perseus-Pisces  superclusters are well seen in 
the velocity field of galaxies, traced
by CosmisFlows data. Among other structures, the Shapley
supercluster stands out also in the velocity field of
the nearby Universe using the newest CosmicFlows data \citep{2023A&A...678A.176D, 2025arXiv250201308C}.

{\it  {The Vela  supercluster}} in the direction of the zone of avoidance at redshift 
$z \approx 0.06$ is a massive supercluster, discovered by \citep{2017MNRAS.466L..29K}. They
showed that this supercluster  consists of two structures with rich clusters,
observed by \citep{2023MNRAS.522.2223H}.

{\it  {The Horologium-Reticulum supercluster}} in the Southern sky at a redshift approximately
$z \approx 0.08$ with its 27 rich (Abell) clusters in the \citep{1994MNRAS.269..301E} catalogue
has been studied in detail by \citep{2005AJ....130..957F}. They showed that galaxy clusters in this supercluster
form two components in redshift space. With a length over 100~\Mpc\, the
Horologium-Reticulum supercluster is one of the largest superclusters in the Southern sky. The authors of
\citep{2005AJ....130..957F, 2010ASPC..423...81F} estimate that the lower limit of the mass of the Horologium-Reticulum supercluster
is $M \approx 5\times~10^{16}~h^{-1}M_\odot$, comparable to that of the Shapley supercluster.

One of the richest superclusters in the Northern sky in the nearby Universe is 
{\it  {the Corona Borealis supercluster,}} 
 at redshift $z = 0.07$
with a mass of $M \approx 1.3\times~10^{16}M_\odot$
\citep{2021A&A...649A..51E} (and references therein). 
The core region with three very rich galaxy clusters (A2065, A2061/A2067, and A2089) is at turnaround,
and these clusters will form the second most massive collapsing structure 
in the nearby Universe in the distant future, with the mass and radius
approximately $M \approx 4.7\times~10^{15}M_\odot$ and
$R \approx 12.5$~\Mpc\  \citep{1988AJ.....95..267P, 1998ApJ...492...45S, 2014MNRAS.441.1601P, 2021A&A...649A..51E,
2024PASA...41...78Z}.

{\it The Aquarius supercluster} at redshift $z = 0.086 - 0.112$ actually consists of two superclusters,
one with five galaxy clusters at redshift $z = 0.086$, and another,
containing 14 galaxy clusters, at redshift $z = 0.112$. The second supercluster may by connected with
a galaxy filament extending up to redshift $z \approx 0.14$ \citep{1999ApJ...520..491B,
2002AJ....123.1200C, 2004AJ....128.2642C}.

Another interesting supercluster in the nearby Universe within the  redshift range $z < 0.1$
is {\it  the supercluster SCl~A2142} at redshift $z = 0.09$ \citep{2015A&A...580A..69E}. 
In comparison with the Shapley and the Corona Borealis
superclusters, SCl~A2142 is poorer, containing one very rich galaxy cluster only.
The central region of the cluster A2142 is mostly populated
with galaxies with very old stellar populations. Such galaxies are also abundant in
the low-density region surrounding this supercluster \citep{2020A&A...641A.172E}.
The overall structure of this supercluster is  interesting, as it contains an
almost spherical, collapsing main body with the elongated central cluster A2142 aligned along the
supercluster axis, and a long, almost straight filament-like tail which extends
farther away from the supercluster and forms 
the longest straight structure in the cosmic web detected so far, with a length
of approximately \mbox{75~\Mpc\ \citep{2020A&A...641A.172E}}.
In total, the supercluster SCl~A2142 has 6--7 filaments connected to
it. This is similar to the connectivity of the Corona Borealis and the Shapley
superclusters \citep{2021A&A...649A..51E, 2024A&A...689A.332A}. 

On ancient maps, the distant, unknown regions were sometimes marked as ``here live the dragons''.
In the cosmic web, we can indicate where the ``dragons''  actually live. Namely,
ref. \citep{1997A&AS..123..119E} determined a rich supercluster at redshift $z \approx 0.1$ in the 
direction of the Draco constellation, which they named {\it  {the Draco supercluster}}. In their list, the Draco
supercluster embeds 16 rich clusters. We may say that the Draco supercluster
guards the borders of the nearby Universe, as the redshift $z \approx 0.1$ is often considered
an approximate limit for the nearby Universe. 

A rich supercluster at slightly higher redshift in the Southern sky is {\it  {the
Sculptor supercluster}} at redshift $z \approx 0.113$ with 22 Abell clusters in it
\citep{2001AJ....122.2222E}. It has an approximate length of $150$~\Mpc, 
and the lower limit of its mass is $M \approx 2\times 10^{16}~h^{-1}M_\odot$
\citep{1998A&A...336...35J, 2008ApJ...685...83E}. 

Moving to higher redshifts, {\it  {the Abell 901/902 supercluster}} at redshift $z = 0.165$
is another example of
a supercluster consisting of two galaxy clusters connected by a filament with galaxy groups
\citep{2008MNRAS.385.1431H}. It stands out among other superclusters
because of the studies of the dark matter distribution in it using a weak lensing analysis
of the structures within this supercluster \citep{2008MNRAS.385.1431H}. 
In this paper, the authors detected four main structures
in the supercluster, which followed the distribution of matter in clusters and 
between them. Most importantly, they concluded that visible galaxies and clusters
in the supercluster trace the distribution of
dark matter well, with no need for the extra 
dark matter which does not follow the distribution of visible matter.

Several very rich superclusters were described for the first time
by \citep{2023ApJ...958...62S} in their supercluster catalogue, which covers the redshift range $0.05 \leq z \leq 0.42$.
The richest among them is the supercluster at redshift $z = 0.25$, centred at the rich cluster
A1835. The authors of the catalogue named this supercluster 
as {\it  {the Einasto supercluster}} in honour of Prof. Jaan Einasto,
who is one of the discoverers of the cosmic web. Jaan Einasto has studied galaxy superclusters for decades,
and is actively contributing to the field now \citep{Einasto:2024aa}. Ref. \citep{2023ApJ...958...62S} estimate
that the mass and extent of the Einasto supercluster are $M \approx 2.6\times 10^{16}M_\odot$ and  $110$~\Mpc.

Another remarkable supercluster in the \citep{2023ApJ...958...62S} catalogue
is {\it  {the Saraswati supercluster}} at redshift $z = 0.28$,
discovered by \citep{2017ApJ...844...25B}. 
In total, this supercluster embeds
51 galaxy groups and clusters, the most massive among them being cluster A2631.
The total mass of the Saraswati supercluster
is of the order of $M \approx 2.6\times 10^{16}M_\odot$, and the length
of it is approximately 200~\Mpc~\citep{2017ApJ...844...25B, 2023ApJ...958...62S}. 
The Saraswati supercluster has a collapsing 
core, which is comparable to the collapsing cores of superclusters in the nearby Universe,
with the mass and radius of the order of  $M \approx 4\times 10^{15}M_\odot$
and $R \approx 20$~Mpc.

The richest supercluster detected so far at redshifts $z \approx 0.5 - 0.6$ is {\it  {the King Ghidorah
supercluster}}, discovered by \citep{2023MNRAS.519L..45S}. The King Ghidorah
supercluster embeds several density enhancements with more than ten galaxy clusters, connected by 
filaments. 
Ref. \citep{2023MNRAS.519L..45S} estimated that its mass is approximately $M \approx 1.1\times~10^{16}M_\odot$.

At redshifts $z \approx 0.9 $, several superclusters have been discovered
\citep{2008ApJ...684..933G, 2013ApJ...768..104F, 2016ApJ...821L..10K, 2019PASJ...71..112H}. The most massive
among them is the supercluster at redshift  $z = 0.9$, with a 
mass  of the order of  \mbox{$M$ = 0.1--$5 \times 10^{15}M_\odot$}
and size $\approx 15$~Mpc \citep{2016ApJ...821L..10K}.
At still higher redshifts, the richest proto-supercluster discovered so far is {\it  {the Hyperion
proto-supercluster}} at redshift $z = 2.45$ \citep{2018A&A...619A..49C}. The authors estimated
that its mass is of the order of $M$  = 0.1--$2.7 \times 10^{14}M_\odot$ and its maximal
extent of the order of $\approx 150$~Mpc. At redshift $ z \approx 3.3$, a proto-supercluster,
nicknamed {\it  {Elent{\'a}ri}}, was discovered in the COSMOS field \citep{2023MNRAS.526L..56F}.

\section{Great Walls as Complexes of Rich Superclusters}
\label{sect:comp}

\begin{figure}[H]
%\isPreprints{\centering}{} % Only used for preprints
\includegraphics[width=12 cm]{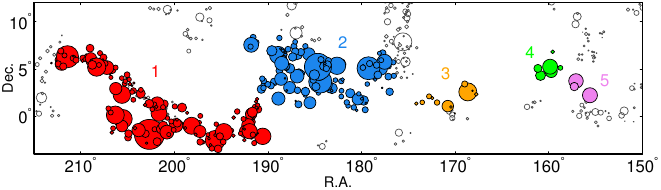}
\caption{Distribution  of galaxy groups  in the Sloan Great Wall superclusters 
in the sky plane in the redshift range $0.04 < z < 0.12$.
Different colours refer to individual superclusters in the SGW;
superclusters 1 and 2 are the two richest SGW superclusters.
Grey symbols show galaxy groups that are not members of the SGW.
Symbol sizes are proportional to the size of groups in the sky plane.
Figure by ME.
\label{sgw}}
\end{figure}

At the largest overdensities of the cosmic web, very rich superclusters form complexes 
in which several rich (and poor) superclusters are almost
connected. In the nearby Universe, the richest supercluster complex is {\it  {the Sloan Great Wall}}
(SGW) 
\citep{2005ApJ...624..463G, 2011ApJ...736...51E, 2012ApJ...759L...7P}.
Its richest superclusters were mapped already in early supercluster catalogues
\citep{1994MNRAS.269..301E, 1997A&AS..123..119E}. 
Ref. \citep{1998A&A...336...35J} noticed that the richest supercluster
among them is one of the flattest, interpreting this as an effect of peculiar motions
of galaxies and galaxy groups toward the supercluster axis, perpendicular to the line-of-sight. 
The definitions of the Sloan Great Wall differ in various studies. Ref. \citep{2012ApJ...759L...7P}
consider this agglomeration of galaxy clusters as one huge, approximately 200~Mpc
long structure. In \citep{2012A&A...539A..80L, 2023ApJ...958...62S} supercluster
catalogues, the Sloan Great Wall consists of several rich and poor superclusters.
In Figure~\ref{sgw}, we show the sky distribution of galaxy groups and
clusters in the SGW region. 

In the extensive study of the morphology and galaxy content 
of the two richest superclusters of the Sloan Great Wall, ref. \citep{2011ApJ...736...51E} showed that these superclusters
have different morphology and galaxy content. The richest supercluster in the SGW
is of multibranching filament morphology, while the second richest supercluster
has multispider morphology; on average, the richest supercluster in the SGW contains a
higher percentage of red, quiescent galaxies than the second richest supercluster.
Considering these differences, ref. \citep{2011ApJ...736...51E} concluded that the formation history and evolution of
individual superclusters in the SGW have been different, and the SGW is not a genuine physical structure
but rather an assembly of very rich galaxy systems.
The total length  of the SGW is approximately $\approx$ 200~Mpc,
and its mass is $M \approx 2\times 10^{16}~h^{-1}M_\odot$.
The richest superclusters in the SGW contain several high-density cores that will
collapse in the distant future, and the SGW may split into several superclusters 
\citep{2016A&A...595A..70E}.

The studies that analysed whether the presence of such a rich complex of superclusters 
in the nearby Universe is in agreement with the standard cosmological model
still leave this as an open question. While some studies found that
the presence of both the Shapley supercluster and the SGW in our cosmic neighbourhood
is hardly compatible with the standard $\Lambda$CDM model,
other studies showed that such systems are present in very large
simulations \citep{2011MNRAS.417.2938S, 2012ApJ...759L...7P}.
Moreover, ref. \citep{2012ApJ...759L...7P} even predicted that there might be another, even
bigger and more massive supercluster system within redshift $z < 0.8$.

\begin{figure}[H]
%\isPreprints{\centering}{} % Only used for preprints
\includegraphics[width=12 cm]{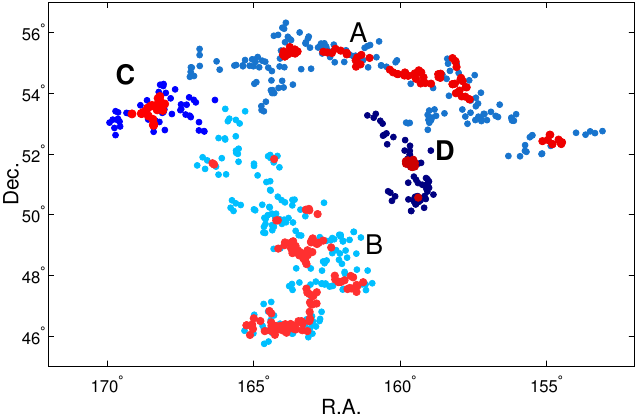}
\caption{Distribution of galaxies of the  BGW superclusters in the sky plane. 
Red dots denote galaxies in the HDCs of each supercluster, and 
blue dots show galaxies in the outskirts. Different shades of red and blue 
correspond to different BGW superclusters, as shown in the figure.
Adapted from \citep{2021A&A...649A..51E}.  
\label{bgwradec}}
\end{figure}

In 2016,  a huge supercluster
complex at redshift $z = 0.47$, consisting of four rich and very rich superclusters,
 was discovered  \citep{2016A&A...588L...4L}.
This supercluster complex obtained a nickname {\it  {the BOSS Great Wall}} (BGW).
The sky distribution of galaxies in the HDCs and outskirts of the four BGW superclusters
is plotted in Figure~\ref{bgwradec}.

The richest supercluster in the BGW alone, BGW A, has a length approximately equal to the
total length of the SGW. The total mass of the BGW superclusters
exceeds the mass of the SGW at least twice
\citep{2016A&A...595A..70E}. In comparison with the nearby rich superclusters,
the BGW superclusters are more elongated, and the reason why there are no such elongated superclusters
in our cosmic ne\textls[-15]{ighbourhood is not yet clear \citep{2016A&A...588L...4L, 2016A&A...595A..70E}. 
The Perseus-Pisces supercluster is also very elongated but not as rich, massive, and large as the BGW superclu}sters.

An analysis of the structure and mass distribution of the BGW superclusters
showed that these superclusters contain altogether eight high-density cores
\citep{2022A&A...666A..52E}. These 
HDCs may form separate superclusters in the distant future, as discussed in Section~\ref{sect:evol}. 
The most massive of these HDCs
has a mass similar to that of the collapsing core in the Corona Borealis supercluster
in the nearby Universe,  $M \approx 3.3\times 10^{15}~h^{-1}M_\odot$.
This may weaken, but does not remove  the tension with the $\Lambda$CDM model, 
which does not predict a large number of very rich and large
superclusters in our nearby cosmic neighbourhood. 
It is also an open question whether the BGW fits the prediction by \citep{2012ApJ...759L...7P},
as superclusters in this study and in~\citep{2016A&A...588L...4L} have been defined in a different way.

Recently, a huge structure was detected, using data on X-ray clusters, and nicknamed
as  {\it  {the Quipu superstructure}} \citep{2025arXiv250119236B}.
Cluster distribution in this structure is rather sparse, and these were linked 
together applying the Friend-of-Friend method with a large linking length.
Therefore, the authors call it a ``superstructure'', not a supercluster.
The Quipu superstructure embeds several poor superclusters of Abell clusters,
determined in the supercluster catalogue by
\citep{2001AJ....122.2222E}, 
and a long filament of X-ray clusters which connect these superclusters.
This is a good example of how data on different wavelengths complement each other to 
obtain a better view of the Universe.

\section{Supercluster Planes}
\label{sect:planes}

Clusters of galaxies surrounding {\it  {the Local supercluster
plane}} (Supergalactic plane), described by de Vaucouleur 
\citep{1953AJ.....58...30D, 1956VA......2.1584D},
are located along this plane, extending it 
up to the distances of several hundred megaparsecs \citep{1982Natur.300..407Z,
1983IAUS..104..405E, 2022MNRAS.511.5093P, 2023MNRAS.526.4490P}. Radio galaxies up to large distances also follow
the Local supercluster plane, as well as 
X-ray luminous galaxy clusters in our cosmic neighbourhood (see also Figure~\ref{abellx})
\citep{1991AuJPh..44..759S, 2021A&A...651A..15B}. 
The richest superclusters that form  this plane are the Local Supercluster, the Coma supercluster, 
the Pisces-Cetus supercluster, the Perseus-Pisces supercluster, and the 
Hydra-Centaurus \mbox{supercluster \citep{1997A&AS..123..119E, 2021A&A...651A..15B}.} 
The Local supercluster plane separates two local voids, the Northern and the Southern
Local supervoids, 
both having diameter approximately $100$~\Mpc \citep{1983IAUS..104..405E, 1994MNRAS.269..301E, 
1995A&A...301..329L, 2012ApJ...754..131K, 2013AJ....146...69C}.

Moreover, the most luminous early-type galaxies in the nearby Universe 
are also concentrated on this plane, while similarly luminous spiral galaxies
are located farther away from this plane \citep{2022MNRAS.511.5093P}. 
Very luminous early-type galaxies are the brightest galaxies in rich galaxy groups and
clusters, which preferentially lie in superclusters or in filaments in supercluster
neighbourhood \citep{2022A&A...668A..69E, 2024A&A...681A..91E}. In the local Universe, their distribution
follows superclusters in the Local supercluster plane.

Perpendicular to the Local supercluster plane, ref. \citep{1997A&AS..123..119E} 
discovered another plane, {\it  {the Dominant supercluster plane}}, with an extent of at least
$600$~\Mpc.
They listed superclusters which are arranged along the Dominant supercluster plane.
In the Southern sky, the superclusters that form the Dominant
Supercluster Plane are the Aquarius-Cetus (9), the Aquarius~(19),
the Aquarius B (8), the Pisces-Cetus (17), the Horologium-Reticulum (26), the Sculptor (22), 
the Fornax-Eridanus (12), and the
Caelum (11) superclusters. Here, the number in parentheses is the number of rich (Abell)
clusters in these superclusters in the \citep{1997A&AS..123..119E} supercluster catalogue. 
In the Northern sky,  the Corona
Borealis (8), the Bootes (12), the Hercules (12), the Virgo-Coma (16), the
Vela (9), the Leo (9), the Leo A (10), the Leo-Virgo (8), and the Bootes A (10) superclusters 
are arranged along this plane. As the Dominant supercluster plane crosses the Local supercluster
plane, some superclusters belong to both.  Superclusters 
in these planes are assembled in a number of intertwined chains of rich superclusters
\citep{1997A&AS..123..119E}.

It is not yet clear whether the presence of such planes is in agreement with 
our standard cosmological model. For example,
using constrained simulations of the nearby
Universe in the 500 $h^{-1}$ Mpc box \citep{2023A&A...677A.169D} found that 
a plane with the extent of approximately 100~\Mpc\ can be recovered in
simulations. However,  0.28\% of random realisations of simulations only
matched both underdense and overdense regions of the nearby Universe,
and their simulation box was not large enough to test the presence
of a supercluster plane hundreds of megaparsecs in size. Also, they did not
analyse the probability of finding two perpendicular supercluster planes.
%%%
From the other hand, recently detected correlated alignments between galaxy clusters 
at scales up to $200 - 300$ comoving Mpc at redshifts at least $z \simeq 1$ may be a 
signature of such planes \citep{west}.
%
%%%%
Refs. \citep{2002A&A...393....1S, 2021arXiv210602672P, 2023MNRAS.526.4490P, 2024arXiv240518307P} speculate that such planes may be
a signature of long, nearly straight strings.

\section{Regularity in the Distribution of Rich Superclusters }
\label{sect:reg}

In Section~\ref{sect:morph}, the possible relation between the location of superclusters
in the cosmic web and their morphology was discussed. Now, we describe the {\it quasiregular
pattern} discovered in the large-scale distribution of rich superclusters.
The first step in this direction
was the study by \citep{1990Natur.343..726B}, who discovered that in the narrow pencil-beam
survey of galaxies in the direction of the North and South Galactic poles,
the distribution of galaxies had regularly located peaks, which corresponded to the clusters of galaxies.
Next, the study of the 3D distribution of rich galaxy clusters and superclusters 
unveiled an almost regular pattern with the same characteristic distance between superclusters, 
120--140 $h^{-1}$Mpc~\citep{1994MNRAS.269..301E, 1997Natur.385..139E}. 
This pattern is best seen in the 3D distribution
of very rich superclusters of Abell clusters of galaxies across huge  underdense regions with diameters over 100 $h^{-1}$Mpc.
%EE: Please check that the intended meaning has been retained % ME: added "of"
In \citep{1994MNRAS.269..301E}, the authors applied several methods in their study of the large-scale distribution of rich superclusters.
They started with the plots that showed the presence of a quasiregular pattern
of rich superclusters and underdense regions between them. 
Various methods, such as the analysis of the sizes of underdense regions  between the 
rich (Abell) 
clusters and  superclusters,
where underdense regions were defined using two different methods, pencil-beams, and the 3D empty sphere method,
confirmed the presence of quasiregular patterns  in the distribution 
of rich superclusters. 
The authors also analysed the distribution
of the nearest neighbour distances between  rich superclusters. 
All these tests suggested that the distribution
of rich superclusters resembles an almost regular lattice with the  sizes of ``lattice cells''
in the range of 120--140 $h^{-1}$Mpc \citep{1994MNRAS.269..301E}.
Underdense regions between very rich superclusters in this lattice
are not completely empty but contain hierarchical, filamentary structures of faint galaxies
and poor groups (Section~\ref{sect:frac}).
A quasi-periodic structure in the distribution of the SDSS's LOWZ  galaxy dataset 
was detected by \citep{2024MNRAS.527.1813R}.

The analysis of the correlation function and power spectrum of rich clusters
in superclusters demonstrated that the correlation function has a series of maxima 
with an interval of the same scale, 120--140 $h^{-1}$Mpc, and the power spectrum
has a maxima at this scale~\citep{1997Natur.385..139E, 1997MNRAS.289..801E}.
These maxima are not seen in the correlation function of all galaxies,
and also not in the correlation function of all groups and clusters. The reason for that is simple---single galaxies and very poor groups can be found everywhere in the cosmic web, and the distances between
them average out when calculating the correlation function.
If the distribution of rich superclusters in the cosmic web were random, then there were also no such
maxima in the correlation function of clusters in superclusters.  
The maxima in the correlation function appear when rich galaxy clusters, 
as density enhancements, form a quasiregular 3D pattern
\citep{1997MNRAS.289..801E}. 
The presence of the regular
patterns in the distribution of rich clusters was further confirmed by applying
the regularity periodogram \citep{2002A&A...393....1S}.

Intriguingly, a hint of the quasiregular pattern in the distribution of quasar
systems at redshifts $z > 1$ was also found \citep{2014A&A...568A..46E}. The scale of this possible
pattern, 400 $h^{-1}$Mpc, differs from the scale found in the distribution 
of nearby rich superclusters. The question of whether
this patterns is real, or just a visual appearance, is still open.

\begin{figure}[H]
%\isPreprints{\centering}{} % Only used for preprints
\includegraphics[width=10 cm]{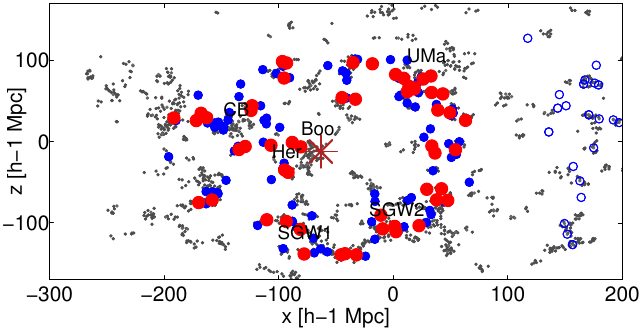}
\caption{Distribution  of galaxy groups  in superclusters 
in Cartesian coordinates in $h^{-1}$~Mpc.
Red-filled circles denote clusters with the number of member galaxies,
$N_{\mathrm{gal}} \geq 50$, and
blue-filled circles denote clusters with $30 \leq N_{\mathrm{gal}} \leq 50$
in a distance interval of $90 \leq D \leq 140$~\Mpc\ from the central cluster in the Bootes supercluster,
A1795 (dark red star).
Blue empty circles denote clusters with $30 \leq N_{\mathrm{gal}} \leq 50$
at a distance interval of $200 \leq D \leq 240$~\Mpc\ from A1795 in another shell.
Grey dots show poor groups with at least four galaxies.
Abbreviations show the names of rich superclusters (Table~\ref{shell}
and \citep{2016A&A...587A.116E}). Adapted from \citep{2016A&A...587A.116E}.
\label{a1795shell}}
\end{figure}

The MAIN galaxy sample of the Sloan Digital Sky Survey covers a part of the pattern of superclusters. 
Three chains of rich superclusters in this region were described in Section~\ref{sect:morph}
\citep{2011A&A...532A...5E}.
Rich superclusters in this sky region  
from the supercluster catalogue by~\citep{2012A&A...539A..80L}, based on the SDSS galaxy data, 
form three chains in an almost round, ``doughnut-shaped'' structure  with a scale of 120--140$h^{-1}$Mpc, 
centred at the cluster A~1795 in the Bootes supercluster, and marked by rich galaxy clusters in superclusters of  
the Sloan Great Wall, the Corona Borealis supercluster, the Ursa Major supercluster, and others 
\citep{2011A&A...532A...5E, 2016A&A...587A.116E}.
\mbox{Figure~\ref{a1795shell}} shows the distribution of 
galaxy groups and clusters  from \citep{2014A&A...566A...1T} group catalogue,
based on the SDSS MAIN galaxy sample, in superclusters centred at cluster A~1795
 in the Bootes supercluster. 
In this figure,  Cartesian coordinates are
defined as in \citet{2012A&A...539A..80L}:
$x = -d \sin\lambda$, $y = d \cos\lambda \cos \eta$, and
$z = d \cos\lambda \sin \eta$, 
where $d$ is the comoving distance, and $\lambda$ and $\eta$ are the SDSS 
survey coordinates. In the SDSS, the survey coordinates form
a spherical coordinate system, 
where $(\eta,\lambda) = (0,90.)$ corresponds to $(R.A.,Dec.) = (275.,0.)$, 
$(\eta,\lambda) = (57.5,0.)$ corresponds to $(R.A.,Dec.) = (0.,90.)$,
and at  $(\eta,\lambda) = (0.,0.)$,  $(R.A.,Dec.) = (185.,32.5)$. 
 In Figure~\ref{a1795shell}, one can see rich clusters
with at least 50 member
galaxies  in superclusters  forming an almost spherical structure with the characteristic size of
approximately 
120--140 $h^{-1}$ Mpc
around cluster A~1795 in the Bootes supercluster.
Clusters in the Hercules supercluster, almost projected to the Bootes supercluster, are located
at the closest to us edge of the structure \citep{2016A&A...587A.116E}. 
At the farthest edge, this structure is bordered by a giant underdense region, described earlier by
\citep{2011A&A...532A...5E, 2012ApJ...759L...7P}, and noted also in \citep{2016A&A...587A.116E}.

%%%%%%%%%%%%%%%%%%%%%%%%%%%%

\begin{table}[H]\vspace{-9pt}
\caption{Data of superclusters.\label{shell}}
	\begin{adjustwidth}{-\extralength}{0cm}
		\begin{tabularx}{\fulllength}{CcCcCcC}
%\begin{tabularx}{\textwidth}{CCC}
\toprule
\textbf{(1)}&\textbf{(2)}&\textbf{(3)}&\textbf{(4)}&\textbf{(5)}& \textbf{(6)}&\textbf{(7)}\\      
\midrule
\textbf{ID(long)} & \textbf{$Dist.$} & \textbf{$L_{\mathrm{tot}}$} & \textbf{$D_{\mathrm{peak}}$} & \textbf{$M_{M/L}$} & \textbf{$M_{gr}$} & \textbf{ID}\\
\midrule
239+027+0091  &  264 & 1809   &  22.2  & 0.65 &  0.74  & A2142             \\
184+003+0077  &  231 & 2919   &  15.0  & 1.01 &  1.06  & SGW2              \\
167+040+0078  &  225 &  751   &  14.6  & 0.32 &  0.43  &  95 *              \\
202-001+0084  &  256 & 5163   &  14.0  & 1.72 &  2.44  & SGW1              \\
230+027+0070  &  215 & 2874   &  11.5  & 1.29 &  0.96  & Corona Borealis   \\
272+054+0071  &  207 & 1618   &   9.2  & 0.72 &  0.19  &  Ursa Major       \\
227+007+0045  &  135 &  379   &   8.7  & 0.16 &  0.27  &  154 *             \\
151+054+0047  &  139 &  465   &   6.8  & 0.20 &        &                   \\
230+008+0030  &  106 & 436    &        & 0.14 &  0.57  &   Hercules1       \\
247+040+0020  &   92 & 527    &        & 0.23 &  0.51  &   Hercules2       \\
\midrule
216+016+0051  &  159 &  284   &   8.5  & 0.12 &  0.27  &   Bootes \\
			\bottomrule
		\end{tabularx}
	\end{adjustwidth}
	\noindent{\footnotesize{The columns are as follows:
1: Long ID of a supercluster AAA+BBB+ZZZZ, 
where AAA is R.A., +/$-$ BBB is Dec. (in degrees), and ZZZZ is 1000$z$ 
(see \citep{2012A&A...539A..80L} for details);
2: the distance of a supercluster, in $h^{-1}$~Mpc;
3: the total weighted luminosity of galaxies in the supercluster, $L_{\mbox{tot}}$,
in $10^{10}~h^{-2} L_\odot$;
4: the luminosity density value at the density maximum (density contrast), $D_{\mbox{peak}}$, 
in units of the mean luminosity density \citep{2012A&A...539A..80L};
5: mass of a supercluster calculated using $M/L = 300$, in units of $10^{16} M_\odot$;
6: mass of a supercluster calculated using the sum of supercluster member groups in the
adaptive supercluster catalogue (\citep{2012A&A...539A..80L}, in units of $10^{16} M_\odot$);
7: ID of the supercluster; (*) denotes the supercluster ID  in  the catalogue by \citet{2001AJ....122.2222E}. 
SGW1 and SGW2 refer
to the two richest superclusters in the Sloan Great Wall, A2142 is the supercluster SCl A2142
\citep{2015A&A...580A..69E}, Hercules1 and Hercules2---the Hercules supercluster, the Corona 
Borealis supercluster (\citep{2021A&A...649A..51E} and references therein), 
the Bootes supercluster, the Ursa 
Major supercluster.
}}
\end{table}

The  structure centred at A~1795 is also seen in the CosmicFlows 4 data, named ``Ho'oleilana'' (``Sent murmurs of awakening'' from the Hawaiian Kumulipo creation chant) 
\citep{2023ApJ...954..169T}. 
Therefore, this structure, as a part of the quasiregular patterns in the distribution
of rich galaxy clusters and superclusters, was determined by different datasets 
(namely, Abell clusters and superclusters, SDSS galaxies, 
groups, superclusters, and CosmicFlows 4 data on peculiar velocities of galaxies and 
galaxy groups)
and different methods,
but all  these show the presence of this pattern, with the characteristic distance between
centres of rich superclusters being approximately 120--140 $h^{-1}$Mpc.

The structure in Figure~\ref{a1795shell} around cluster A~1795 is formed by rich clusters (red circles
in the figure) in the richest superclusters
of the Sloan Great Wall (SGW1 and SGW2), in the Corona Borealis supercluster
(CB), in the Ursa Major supercluster (UMa), and in other superclusters.
In Table~\ref{shell}, we present data on these superclusters,
including supercluster masses. As for all these superclusters,
luminosities are available from \citep{2016A&A...587A.116E}, and the supercluster masses have been calculated
using supercluster luminosities and mass-to-light ratios $M/L = 300$,
as described in Section~\ref{sect:mass}. For comparison, we also provide 
supercluster masses, calculated using the sum of supercluster member group masses.

The mass of the central cluster  in the Bootes supercluster A~1795 is
$M \approx$ 6.6--$11.2 \times10^{14} M_\odot$~\citep{2005ApJ...628..655V, 2014A&A...566A...1T}.
The mass of the  Bootes supercluster itself is the lowest among
superclusters in this shell, $M_{Boo} \approx 0.1 \times10^{16} M_\odot$. 
Masses of superclusters in the walls of this  structure 
are more than one hundred times higher than the mass of the central
cluster, A1795, of the order of $M_{scl} \approx$ 0.2--$3.5 \times10^{16} M_\odot$.
The total mass in superclusters in the shell walls is at least $M_{tot} \approx 25 \times10^{16} M_\odot$.

%%%%%%%%%%%%%%%%%%%%%%%%%%%%

The huge sizes of these structures tell us that the origin of these structures, 
and quasiregular patterns in the supercluster distribution should come from the very early Universe. 
Ref. \citep{2016A&A...587A.116E} concluded that the process
behind these structures is still unknown. Ref. \citep{2023ApJ...954..169T}
interpreted the structure that they named Ho'oleilana as a Baryon Acoustic Oscillation (BAO) shell.
However, in the BAO shells, the central mass is always much higher than the mass in the shell walls 
\citep{2012A&A...542A..34A}.
The analysis of the  masses of superclusters  contradicts 
the interpretation of Ho'oleilana as BAO shell, as proposed in \citep{2023ApJ...954..169T}. 
Ref. \cite{2016A&A...587A.116E} already provided arguments against BAO interpretation. Namely, its radius 
is larger than $\approx 109$ $h^{-1}$ Mpc, the  BAO scale, and structures are wide, 
being in the interval of \mbox{120--140 $h^{-1}$ Mpc. }
In addition, the quasiregular pattern  in the supercluster distribution is traced by the rich
and massive superclusters, and this suggests that its origin is related to the dark
matter distribution, and not to baryon oscillations, as these are much weaker.
The distribution of galaxy clusters and superclusters is given by the initial density perturbation field,
generated during or after the inflation period of the evolution of {\it {the dark matter}}-dominated
Universe. In contrast, baryonic acoustic
oscillations are generated by sound waves in the baryonic matter before
recombination.
It is an open question whether the presence 
of such a quasiregular pattern of rich superclusters can be explained within the $\Lambda$CDM cosmological model,
as also discussed in \citep{2002A&A...393....1S}.

\section{Summary and Outlook}
\label{sect:out}

Summarising the studies of superclusters since the discovery of the cosmic web,
traced by superclusters connected by filaments and separated by voids 
\citep{1980MNRAS.193..353E, 1980Natur.283...47E, 1983ARA&A..21..373O},
we can see a tremendous change. 
First of all, superclusters played a crucial role in the discovery
of the cosmic web: the understanding that  superclusters are connected by filaments
and separated by voids, forming a pattern now called the cosmic web, was introduced for the first time 
by the Tartu group of astronomers in their studies of nearby superclusters
\citep{1978MNRAS.185..357J, 1980MNRAS.193..353E}. This discovery is also
described in \citep{1983ARA&A..21..373O}.
In the first review of superclusters, ref. \citep{1983ARA&A..21..373O}
mentioned that approximately 20 superclusters were known, with a more detailed
description of four nearby superclusters, namely, the Virgo, Coma, Perseus,
and Hercules superclusters. In contrast, the latest supercluster catalogues 
of optical groups and clusters contain data on 662 superclusters in the redshift range
$0.05 \leq z \leq 0.42$ \citep{2023ApJ...958...62S}, and on 633 superclusters 
in the redshift range $0.5 \leq z \leq 1.0$ \citep{2024ApJ...975..200C}. The newest
catalogues of superclusters of X-ray clusters include data on eight superclusters
in the very local Universe within redshift $z < 0.03$ \citep{2021A&A...656A.144B},
and 1338 superclusters in the Western hemisphere up to redshifts 
$z = 0.8$, 520 of them embed at least three member clusters \citep{2024A&A...683A.130L}.
Superclusters in the nearby Universe are also traced using the velocity field
of galaxies up to redshift $z \approx 0.1$ \citep{2017ApJ...845...55P, 
2023A&A...678A.176D, 2025arXiv250201308C}.

The most massive supercluster in the nearby Universe is one of the first 
superclusters discovered, the Shapley supercluster, with a mass of approximately
$M \approx 5\times 10^{16}~h^{-1}M_\odot$. At higher redshifts, several
massive superclusters have been discovered, as the Saraswati, the Einasto, and
the King Ghidorah superclusters, with masses of the same order as the Shapley
superclusters. The extent of the richest superclusters may reach up to $200$~\Mpc.

Superclusters represent a special environment for the formation and evolution of galaxies,
groups, and clusters. They
embed the richest galaxy clusters and the most massive galaxies---the 
brightest galaxies of rich clusters. On average, superclusters host a higher percentage of red,
quiescent galaxies than surrounding low-density regions. X-ray clusters are more frequent
in superclusters. They also affect the radio lobes of giant radio galaxies.
The preferred alignment of galaxies, groups, and clusters along the supercluster axis,
discovered already in the first supercluster studies, has been confirmed by several 
later studies. 
It is interesting to note that in his review on superclusters in 1983, Jan Oort had already
emphasised the importance of the studies of the coevolution of superclusters and 
galaxies and galaxy systems within them
\citep{1983ARA&A..21..373O}. This is one of the important future directions in supercluster
studies also now.
Another direction in the future studies of superclusters is a more detailed analysis
of their dynamical properties, as well as of the fractal properties of superclusters.

Superclusters serve as a cosmological test in many ways. The number, density,
masses, shapes, and other properties 
of superclusters and their high-density cores can be compared with cosmological simulations 
with various cosmologies. The further studies of superclusters will, hopefully,
uncover which are the processes in the early Universe behind the regularity of 
superclusters in the nearby Universe, and whether this regularity can also be detected at
higher redshifts. They will clarify the origin of supercluster planes, now speculated
as being signatures of long strings in the early Universe.

All this has been made possible thanks to the multiwavelength surveys of galaxies, and
from the simulation side, thanks  to the simulations in increasingly larger simulation boxes.
%EE: Please check that the intended meaning has been retained %ME made minor changes
In this respect, cosmology is now facing a new breakthrough epoch in the next 5--10 years.
This will be possible thanks to ongoing and new (full-sky) multiwavelength galaxy surveys,
such as 4MOST, J-PAS, DESI, Euclid, JWST (early galaxies at cosmic dawn), CMB surveys,
SPHEREx, and others, which enable tracing the density and velocity fields of galaxies
up to high redshifts. The combination of various types of data and methods may lead to a better 
definition of superclusters.
The study of superclusters may help to understand the origin of \mbox{Hubble
tension.}

In summary, future deep and wide galaxy surveys will provide the
opportunity to compare the abundance and properties of
superclusters at higher redshifts with the ones in simulations.
Therefore, it will be possible to identify and characterise superclusters up to high redshift so as
to understand supercluster evolution together with the evolution of the cosmic web and galaxies within.
Future studies will hopefully provide us with a better definition of superclusters
that take into account the properties of galaxies and groups in superclusters,
and velocity data with higher precision than is possible at present.
This enables us to compare the properties of galaxies in large,
high-density regions like superclusters and in underdense
regions like voids to understand the factors affecting their
growth and evolution. The current era of deep multiwavelength
large sky surveys provides us with the perfect opportunity.

%%%%%%%%%%%%%%%%%%%%%%%%%%%%%%%%%%%%%%%%%%%%%%%%%%%%

%%%%%%%%%%%%%%%%%%%%%%%%%%%%%%%%%%%%%%%%%%

\funding{This  work was supported by the Estonian Ministry of Education and Research (grant TK202,
 “Foundations of the Universe''), 
Estonian Research Council grant PRG1006, by Estonian Research Council grant PRG2172, and the 
European Union's Horizon Europe research and innovation programme 
(EXCOSM, grant No. 101159513). }

\dataavailability{No new data were created, except supercluster masses in Table~\ref{shell}.}

\acknowledgments{{I thank} %MDPI: Please ensure that all individuals included in this section have consented to the acknowledgement.
 the referees for inspiring reports which helped me to improve the paper.
This review is inspired by  discussions with many people.
First of all, I thank Jaan Einasto, Shishir Sankhyayan,
Pekka Hein{\"a}m{\"a}ki, and Lauri Juhan Liivam{\"a}gi for carefully reading the manuscript and 
for many useful suggestions. My special thanks to Lauri Juhan Liivam{\"a}gi
for providing Figure~\ref{denlevels}, to Pekka Heinämäki
for Figure~\ref{lsig}, to Jaan Einasto for Figures~\ref{sdssslice} and \ref{fill},
to Helene Courtois and Alexandra Dupuy for Figure~\ref{baos}, Nabila Aghanim
for Figure~\ref{shapsz}, 
and to Shishir Sankhyayan for Figure~\ref{phase}.
\mbox{Figures~\ref{k12}, \ref{dfrac}, \ref{denevol}, \ref{a2142sky}, \ref{a2142zones},
\ref{bgwsky}, \ref{cores}, \ref{d8dfil}, \ref{frac}, \ref{bgwradec}, and \ref{a1795shell}} 
are reproduced with permission from Astronomy \& Astrophysics, \copyright ESO,
and Figures~\ref{mfabell} and \ref{phase} are reproduced with permission from the AAS.
%I thank Helene Courtois, Alexandra Dupuy, Nabila Aghanim, Mirt Gramann, Mathieu Colless,
%Hans B{\"o}hringer, Gayoung Chon, Toni Tuominen, Yen-Ting Lin, Changbom Park,
%Rogerio Monteiro-Oliveira, Renyue Cen, and 
I thank my colleagues at Tartu Observatory and elsewhere
for fruitful and interesting discussions. 
This review is also related to COST Action CA21136 “Addressing observational 
tensions in cosmology with systematics and fundamental physics (CosmoVerse)”, 
supported by COST (European Cooperation in Science and Technology). 
}

\conflictsofinterest{The author declares no conflicts of interest. 
The funders had no role in the design of the study; 
in the collection, analyses, or interpretation of data; in the writing of the manuscript; 
or in the decision to publish the results.}

%%%%%%%%%%%%%%%%%%%%%%%%%%%%%%%%%%%%%%%%%%
%% Optional

%\abbreviations{Abbreviations}{
%The following abbreviations are used in this manuscript:\\
%
%\noindent 
%\begin{tabular}{@{}ll}
%HDC & High-density core\\
%%DOAJ & Directory of open access journals\\
%%TLA & Three letter acronym\\
%%LD & Linear dichroism
%\end{tabular}
%}

%%%%%%%%%%%%%%%%%%%%%%%%%%%%%%%%%%%%%%%%%%
%% Optional
%appendixtitles{no} % Leave argument "no" if all appendix headings stay EMPTY (then no dot is printed after "Appendix A"). If the appendix sections contain a heading then change the argument to "yes".

% Appendix: supercluster catalogues

%%%%%%%%%%%%%%%%%%%%%%%%%%%%%%%%%%%%%%%
\begin{adjustwidth}{-\extralength}{0cm}

\printendnotes[custom]

\reftitle{References}

%\bibliography{scl.bib}

\PublishersNote{}
\end{adjustwidth}

% If authors have biography, please use the format below
%\section*{Short Biography of Authors}
%\bio
%{\raisebox{-0.35cm}{\includegraphics[width=3.5cm,height=5.3cm,clip,keepaspectratio]{Definitions/author1.pdf}}}
%{\textbf{Firstname Lastname} Biography of first author}
%
%\bio
%{\raisebox{-0.35cm}{\includegraphics[width=3.5cm,height=5.3cm,clip,keepaspectratio]{Definitions/author2.jpg}}}
%{\textbf{Firstname Lastname} Biography of second author}

% For the MDPI journals use author-date citation, please follow the formatting guidelines on http://www.mdpi.com/authors/references
% To cite two works by the same author: \citeauthor{ref-journal-1a} (\citeyear{ref-journal-1a}, \citeyear{ref-journal-1b}). This produces: Whittaker (1967, 1975)
% To cite two works by the same author with specific pages: \citeauthor{ref-journal-3a} (\citeyear{ref-journal-3a}, p. 328; \citeyear{ref-journal-3b}, p.475). This produces: Wong (1999, p. 328; 2000, p. 475)

%%%%%%%%%%%%%%%%%%%%%%%%%%%%%%%%%%%%%%%%%%
%%%%%%%%%%%%%%%%%%%%%%%%%%%%%%%%%%%%%%%%%%

%\isPreprints{}{% This command is only used for ``preprints''.
%\end{adjustwidth}
\end{document}